\documentclass[a4paper,fleqn,usenatbib]{mnras}

\usepackage[T1]{fontenc}
\usepackage{ae,aecompl}

\usepackage{graphicx}	% Including figure files
\usepackage{amsmath}	% Advanced maths commands
\usepackage{amssymb}	% Extra maths symbols
\usepackage{subfig}

\title[Poincar\'e surfaces of section around a 3-D irregular body]{Poincar\'e surfaces of section around a 3-D irregular body: The case of asteroid 4179 Toutatis}

\author[G. Borderes-Motta; O. C. Winter]{%
  G. Borderes-Motta$^{1}$\thanks{E-mail: gabriel\underline{ }borderes@yahoo.com.br (GBM)} and
  O. C. Winter$^{1}$\thanks{E-mail:  ocwinter@gmail.com (OCW)} 
  \\
  $^{1}$ Grupo de Din\^amica Orbital e Planetologia, S\~ao Paulo State University, 12516-410 Guaratinguet\'{a}, 
  S\~{a}o Paulo, Brazil}

\date{Accepted XXX. Received YYY; in original form ZZZ}

\pubyear{2015}

\begin{document}
\label{firstpage}
\pagerange{\pageref{firstpage}--\pageref{lastpage}}
\maketitle

\begin{abstract}
In general, small bodies of the solar system, e.g., asteroids and comets, have a very irregular shape. This feature affects significantly the gravitational potential around these irregular bodies, which hinders dynamical studies. The Poincar\'e surface of section technique is often used to look for stable and chaotic regions in two-dimensional dynamic cases. In this work, we show that this tool can be useful for exploring the surroundings of irregular bodies such as the asteroid 4179 Toutatis. Considering a rotating system with a particle, under the effect of the gravitational field computed three-dimensionally, we define a plane in the phase space to build the Poincar\'e surface of sections. Despite the extra dimension, the sections created allow us to find trajectories and classify their stabilities. Thus, we have also been able to map stable and chaotic regions, as well as to find correlations between those regions and the contribution of the third dimension of the system to the trajectory dynamics as well. As examples, we show details of periodic(resonant or not) and quasi-periodic trajectories.
\end{abstract}

\begin{keywords}
minor planets, asteroids: individual: 4179 Toutatis -- methods: numerical --  celestial mechanics.
\end{keywords}

%%%%%%%%%%%%%%%%%%%%%%%%%%%%%%%%%%%%%%%%%%%%%%%%%%

%%%%%%%%%%%%%%%%% BODY OF PAPER %%%%%%%%%%%%%%%%%%

\section{Introduction}

Many bodies of the solar system do not have enough mass to be in hydrostatic equilibrium. Such condition allows the bodies to have an irregular and asymmetrical shape, as it happens with comets and asteroids. These astronomical objects arouse the interest of the scientific and civil community due to information they provide about the beginning of the solar system, the danger that they may represent to Earth \citep{Shustov} and the mining missions that can be performed in these bodies\citep{Abell}. Besides this growing interest in those astronomical bodies, the number of space missions to those bodies are growing as well. Some examples of those space missions are shown as follow.

NASA's mission NEAR-Schoemaker\citep{NShoemaker}, JAXA's Hayabusa\citep{Haya2-b} and ESA's ROSETTA\citep{rosetta} visited asteroids  433 Eros, 25143 Itokawa and comet 67P/Churyumov-Gerasimenko, respectively and they are examples of successful missions sent to irregular bodies. Currently, the missions JAXA's Hayabusa-2\citep{Haya2-b} and NASA's OSIRES-Rex\citep{Haya2-b} aim to explore the asteroids 162173 1999 JU3 and 101995 Bennu respectively. Some daring maneuvers were performed in some of the missions quoted and will be repeated in the others. NEAR-Schoemaker spacecraft was the first one to touch the surface of a small body (433 Eros). The Hayabusa spacecraft accomplished a touch-down on the surface of 25143 Itokawa. During this maneuver, the spacecraft collected material from the surface of the asteroid, which was sent to Earth\citep{Haya2-b}. Hayabusa-2 \citep{Haya2-b} and OSIRES-Rex \citep{Osires-b} intend to collect material from their respective asteroids. Hayabusa-2 \citep{Haya2-b} and ROSETTA \citep{rosetta} even use modules for studies in situ.

For the missions to be successful, the spacecrafts must orbit the targets and collect data using their tools, as mentioned previously. For this reason accurate studies on orbital dynamics around irregular bodies are needed. Therefore, many works  have been made enriching the knowledge concerning the structure of gravitational potential of irregular bodies. \citet{Frouard} used two numerical models to study the short and long-term evolution of the asteroid system 87 Sylvia, they showed the deeply stable zone, fast and secular chaotic regions of the system. \citet{Araujo2012} used N-body simulations and calculated stable regions around triple asteroid 2001 SN263. They found the size and the location of the stabilities zones within the system. \citet{2015JIANG} used the periodic orbits to study the stable region in the potential field of the primary body of triple asteroid 216 Kleopatra. In the present work, we show the usefulness of one of those methods that allows to explore regions around irregular and asymmetric bodies and determine the size and location of the stable and unstable regions.

In order to explore the dynamical region around an irregularly shaped body, we adopt the Poincar\'e surface of section technique. It brings the possibility of identifying the location and size of the regular and chaotic regions, as well as providing information about resonances, periodic orbits, and quasi-periodic orbits. This technique has been widely and successfully applied to orbital dynamics. More than half century ago, a significant number of numerical studies of the restricted three-body problem via Poincar\'e surface of section started \citep{henon-a,henon-b,henon-c,henon-d,henon-e,jefferys}. A more comprehensive research for a specific mass ratio followed those studies a couple of decades ago \citep{atlas1,atlas2}. Using the restricted three-body problem for a Sun-Jupiter system, \citet{WM97a,WM97b} analyzed first-order resonances and libration regions. \citet{w00} applied the Poincar\'e surface of section technique to the Earth-Moon system, where stability regions and periodic orbits were found, and the maximum libration amplitude of the quasi-periodic orbits around a family of periodic orbits was computed. In all the mentioned works, the Poincar\'e surface of section technique was applied to the two-dimensional problem; the planar, circular restricted three-body problem. 

However, we are interested in applying the technique to irregular and non-symmetric bodies, a three-dimensional problem. \citet{Sh96} used the Poincar\'e map in order to find a periodic orbit around the asteroid  4769 Cast\'alia, whose gravitational potential is irregular and three-dimensional. In the context of the application of Poincar\'e surface of section technique to a single body, \citet{BE05} considered a circular ring to model a gravitational potential and find several families of periodic orbits, as well as dynamic structures around the ring. Since the ring is a planar object, the problem is also planar. Another planar case was used in \citet{silva09}. The authors analyzed the region around planar shapes (a square and a triangular plate) via Poincar\'e surface of section and determined stable and unstable regions. \citet{Naj11} used the Poincar\'e surface of sections around an inhomogeneous straight segment. Such segment was set out perpendicular to the rotating system defined. The gravitational potential generated by this segment is two-dimensional, since the non-homogenity of the segment density is a function that makes the gravitational potential symmetric. \citet{liu11} used the polyhedron model to compute the gravitational potential of a three-dimensional cube, thus, enabling the mapping by the Poincar\'e surface of section at the cube vicinity and periodic orbits around the cube were found. All the mentioned works treat two-dimensional or three-dimensional cases. However, whenever it was a three-dimensional case it had symmetry, which made it turn into a two-dimensional case as well. Differently from then, \citet{Sh96} took into account the third dimension in a non-symmetric system, focusing only on determining periodic orbits. Stable and unstable regions were not explored at all.

In order to broadly and systematically explore the stable and chaotic regions around irregular and asymmetric bodies, we use the Poincar\'e surface of section technique. Our aim is to show that, even with more degrees of freedom, the known structures of Poincar\'e surfaces of section are not in general destroyed. In this way, the technique is not restricted to look for periodic orbits, but also to allow the mapping of all the dynamic structures around the body.

In the next two sections we introduce the main tools used in the present work; the mascons model and the Poincar\'e surface of section technique. In section \ref{sec:results} we present and discuss the results for a study about the asteroid 4179 Toutatis. Stable regions, periodic orbits, quasi-periodic orbits and chaotic regions around this asteroid are identified and analyzed. In the last section, the final comments and general observations are presented.
%%%%%%%%%%%%%%%%%%%%%%%%%%%%%%%%%%%%%%%%%%%%%%%%%%
%%%%%%%%%%%%%%%%%%%%%%%%%%%%%%%%%%%%%%%%%%%%%%%%%%
%%%%%%%%%%%%%%%%%%%%%%%%%%%%%%%%%%%%%%%%%%%%%%%%%%
\section{Shape model: mascons}
\label{sec:masc}

The computation of gravitational potential of irregular and small bodies composes the foundation for a dynamical study around these bodies. This physical amount can be represented as an expansion by harmonic series for any body \citep{Mac36}. The gravitational potential is better understood and represented, in terms of accuracy, when its harmonic series have a high number of gravitational coefficients. Another method to evaluate the gravitational field, via the dimensions of the body, is given by Ivory's approach. Which consists of fitting a triaxial ellipsoid whose dimensions resemble the body \citep{La82,Ke54}. 

The polyhedron shape model, a tool that has been developed to model irregular bodies, has been helping to compute a gravitational potential closer to a real one. It provides, with a high level of accuracy, the shape of an observed body \citep{radar}. The shape given by this model is usually a polyhedron of triangular faces that fits the real irregular surface of the body. In order to simplify the computational procedures, it is recommended to decompose the polyhedron into other geometric shapes, like tetrahedrons. With these tetrahedrons, at least two approaches are possible. The first approach is called polyhedron model, and it was developed by \citet{W94}. It is based on a volumetric integration of tetrahedrons in order to obtain the gravitational potential. The other approach was firstly applied in \citet{G96} and then called ``mascons'' model (mass concentration) \citep{WS96,rossi99}. Such model is based on filling the polyhedron volume with massive points in order to reproduce the mass distribution of the body. There are, at least, two approaches to distribute these points inside the polyhedron volume. In the first, the points are uniformly distributed in a three-dimensional space \citep{G96}. In the second, the points are systematically set by using the geometric features of the tetrahedrons\citep{venditti}. 

In the present work we adopted the mascons model\citep{G96} whose main advantage is the control that can be exerted on the accuracy and the computational speed. The number of mascons is directly proportional to the accuracy and inversely proportional to the integration speed.

Our mascon model was built starting from a uniform grid of points. This grid has dimensions larger than the original polyhedron\citep{G96} and is centered at the geometric center of the polyhedron. The points of the grid that do not belong to the volume of the polyhedron, are excluded. The limit of the grid is the polyhedron itself, i.e., the grid has the irregular shape of the body. The gravitational potential of the body, which allows to integrate orbits, to evaluate the equilibrium points, and to study other dynamical features, is computed via the contribution of the sum of the massive points.
%%%%%%%%%%%%%%%%%%%%%%%%%%%%%%%%%%%%%%%%%%%%%%%%%%
%%%%%%%%%%%%%%%%%%%%%%%%%%%%%%%%%%%%%%%%%%%%%%%%%%
%%%%%%%%%%%%%%%%%%%%%%%%%%%%%%%%%%%%%%%%%%%%%%%%%%
\section{Poincar\'e surface of section technique}
\label{sec:ssp}

\begin{figure}
\begin{center}
\includegraphics*[width=\columnwidth]{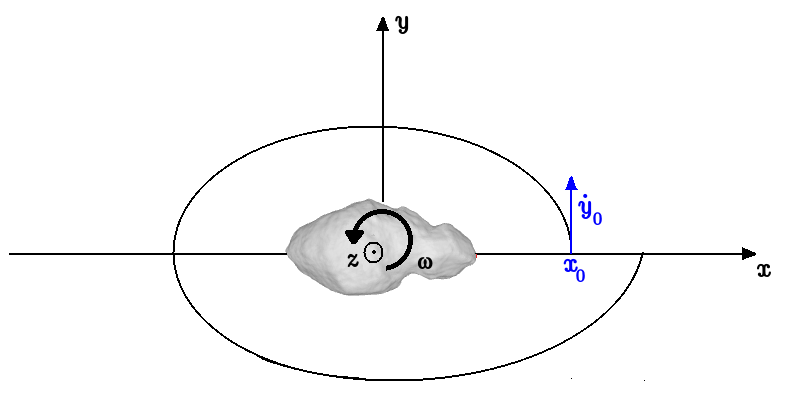}
\end{center}
\caption{Schematic diagram of an orbit around a body fixed at a rotating frame. $x_{0}$ marks the initial position and the blue arrow indicates the initial velocity.}
\label{fig:SSP-1}
\end{figure}

Having defined the mathematical techniques to compute gravitational potential, the equations of motion can be written. These equations are defined in the body-fixed frame (O$xyz$), given in \citep{Sh96}, a frame with uniform rotation that follows the asteroid movement (Fig. \ref{fig:SSP-1}). The $xy$ plane is called, in this work, \textit{rotating plane}.
\begin{equation}
\ddot{x} - 2 \omega \dot{y} = \omega ^{2} x + U_{x}$,$ 
\label{eq:movx}
\end{equation}
\begin{equation}
\label{eq:movy}
\ddot{y} + 2 \omega \dot{x} = \omega ^{2} y + U_{y}$ and$
\end{equation}
\begin{equation}
\label{eq:movz}
\ddot{z} =  U_{z}$,$
\end{equation}
where $U_{x}$, $U_{y}$, and $U_{z}$ stands for the partial derivatives of the gravity potential and $\omega$ is the spin velocity of the asteroid. As mentioned before, the mascons model is used here to compute the gravitational potential, given by 
\begin{equation}
U = U(x,y,z) = - \sum^{N}_{i=0} \frac{G \ m}{r_{i}^{2}},
\end{equation}
where $G$ is the gravitational constant, $N$ is the total number of mascons, $r_{i}$ is the distance from a mascon to the orbiting particle, and $m$ is the mass of each mascon, $m=\frac{M}{N}$ with $M$ being the total mass of the body.

Conserved quantities, such as the Jacobi constant ($C_{j}$), can be useful to analyze the equations of motion. This constant is explicitly computed as \citep {Sh96}:
\begin{equation}
\label{eq-Cj}
C_{j} = \omega ^{2}(x^2+y^2)+2U(x,y,z)-\dot{x}^{2}-\dot{y}^{2}-\dot{z}^{2}.
\end{equation}
The equations (\ref{eq:movx}-\ref{eq:movz}) describe the motion of a massless particle and are numerically integrate via the Burlish-Stoer integrator\citep{BS66}. The Poincar\'e section is set in the plane $y=0$ and the initial conditions are systematically distributed over the $x$ axis. It is defined $y_{0}=z_{0}=\dot{x} _{0}=\dot{z} _{0}\equiv 0$, $\dot{y}_{0}>0$ and $\dot{y}_{0}$ was computed for a fixed value of $C_{j}$ (Eq. \ref{eq-Cj}). During the integration, the conditions of the orbit are saved at every instant of time when the trajectory crosses the section defined by $y=0$ with $\dot{y}>0$. The Newton-Raphson method is used to obtain an error of ($10^{-13}$) from that section. The recorded points are plotted on the phase plane $(x,\dot{x})$ creating the Poincar\'e surface of section.

The method used in this work was built as a planar two-dimensional case, adding the third dimension ($z$) to the particle trajectory. The velocity $\dot{z}$ was set at zero in the initial condition, and during the integration, both the coordinate and the velocity are free to evolve. In order to analyze the results, the movement projected over the \textit{rotating plane} was thought to be uncoupled from the movement on the third dimension. Then, it was studied the influence of $z$ variation of the orbit behavior on the \textit{rotating plane}. In the Poincar\'e surface of section, the most internal orbit of a cluster of stability islands is called, in this paper, \textit{central orbit}. Each \textit{central orbit} is represented by one isolated point in the Poincar\'e surface of section. These points correspond to the periodic orbits in the projection over the \textit{rotating plane}. Thus, the third dimension varies avoiding the repeatability, as it usually occurs in periodic orbits.

Hereinafter, we present a study about the dynamics around the asteroid 4179 Toutatis, which was chosen  due to its low spin velocity. Among the asteroids with necessary data for the mascons approach, the 4179 toutatis has a rotation period of the order of days, whereas most of the other asteroids have it in hours. This period enhances the influence of the irregularity of the gravitational potential over a particle orbit. The asteroid 4179 Toutatis is a NEA and has already had several close encounters with Earth, which provided accurate data on the asteroid. A shape model was built by \citet{Ost95} through radar data taken in December 1992. This shape model represents the asteroid 4179 Toutatis through a polyhedron with $20,000$ vertices and $39,996$ faces \citep{radar}. Its rotational period is $5.376$ days ($T_{ast}$) \citep{Ost95} and the density is $2.5$ $g/cm^{3}$ \citep{Sh98}. We apply the algorithm (Section \ref{sec:masc}) to obtain the mascons and we consider $21,106$ mascons points that reproduce the gravitational potential of the asteroid.

%%%%%%%%%%%%%%%%%%%%%%%%%%%%%%%%%%%%%%%%%%%%%%%%%%
%%%%%%%%%%%%%%%%%%%%%%%%%%%%%%%%%%%%%%%%%%%%%%%%%%
%%%%%%%%%%%%%%%%%%%%%%%%%%%%%%%%%%%%%%%%%%%%%%%%%%
\section{Results}
\label{sec:results}

In our work the regions around the asteroid 4179 Toutatis are explored considering a range of $C_{j}$ values from $1.20$ to $3.0\times 10^{-7}$ $km^{2}/s^{2}$, with $0.15\times 10^{-7}$ $km^{2}/s^{2}$ of interval between them. The unit of $C_{j}$ as $10^{-7}$ $km^{2}/s^{2}$ is adopted by simplicity.

Using $75$ different initial conditions for each $C_{j}$, we simulate the trajectory generated by each condition. The trajectories integration are stopped after $1000$ intersections between the trajectory and the plane $y=0$. Those conditions are distributed along the  $x$ axis from $2.6$ to $10$ $km$ with steps of $0.1$ $km$. These simulations produce the first view of the Poincar\'e surface of section, then, if it is necessary to complete this surface of section, new conditions are simulated.

\begin{figure*}
	\includegraphics[width=14cm,height=7cm]{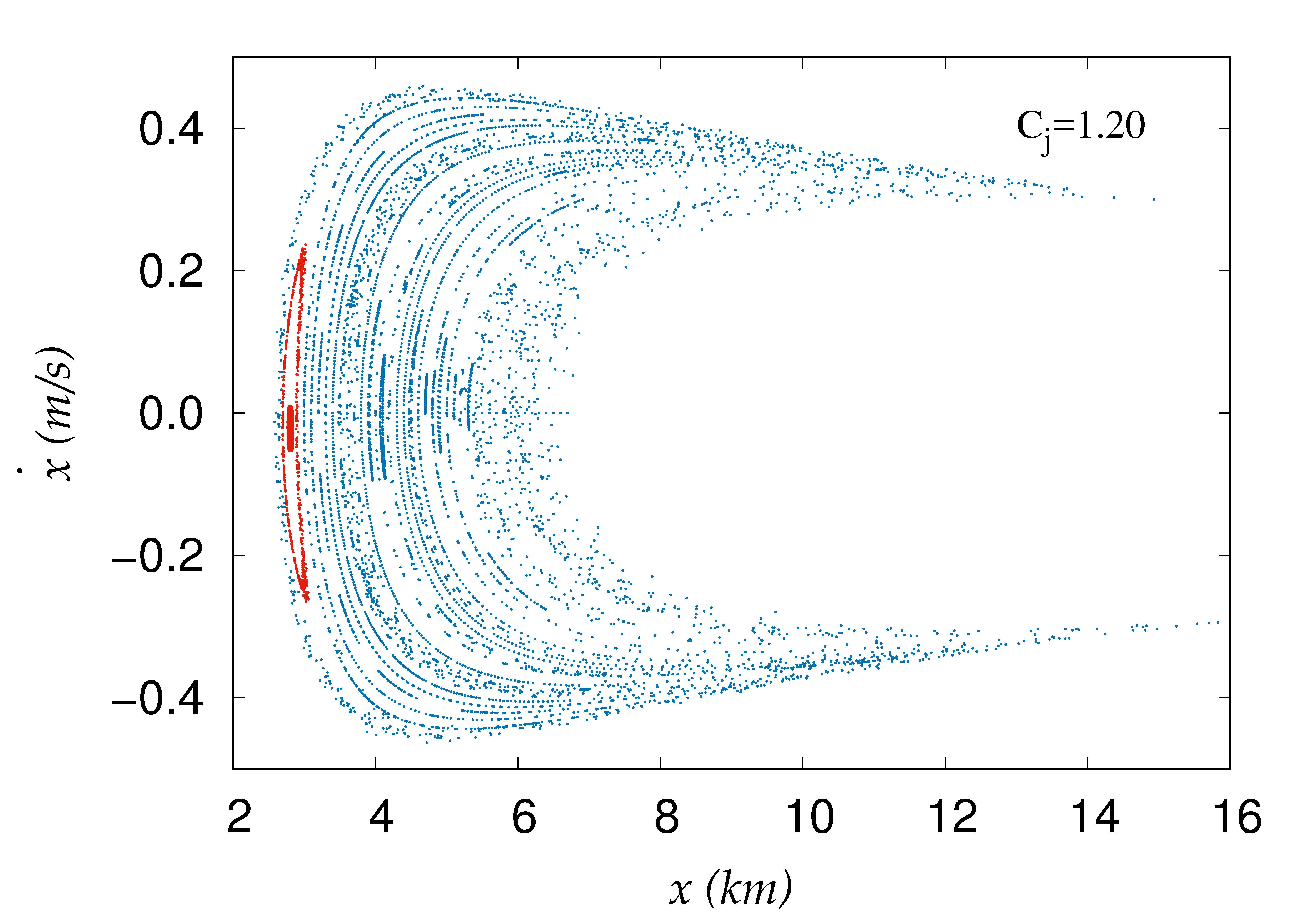}
	\includegraphics[width=14cm,height=7cm]{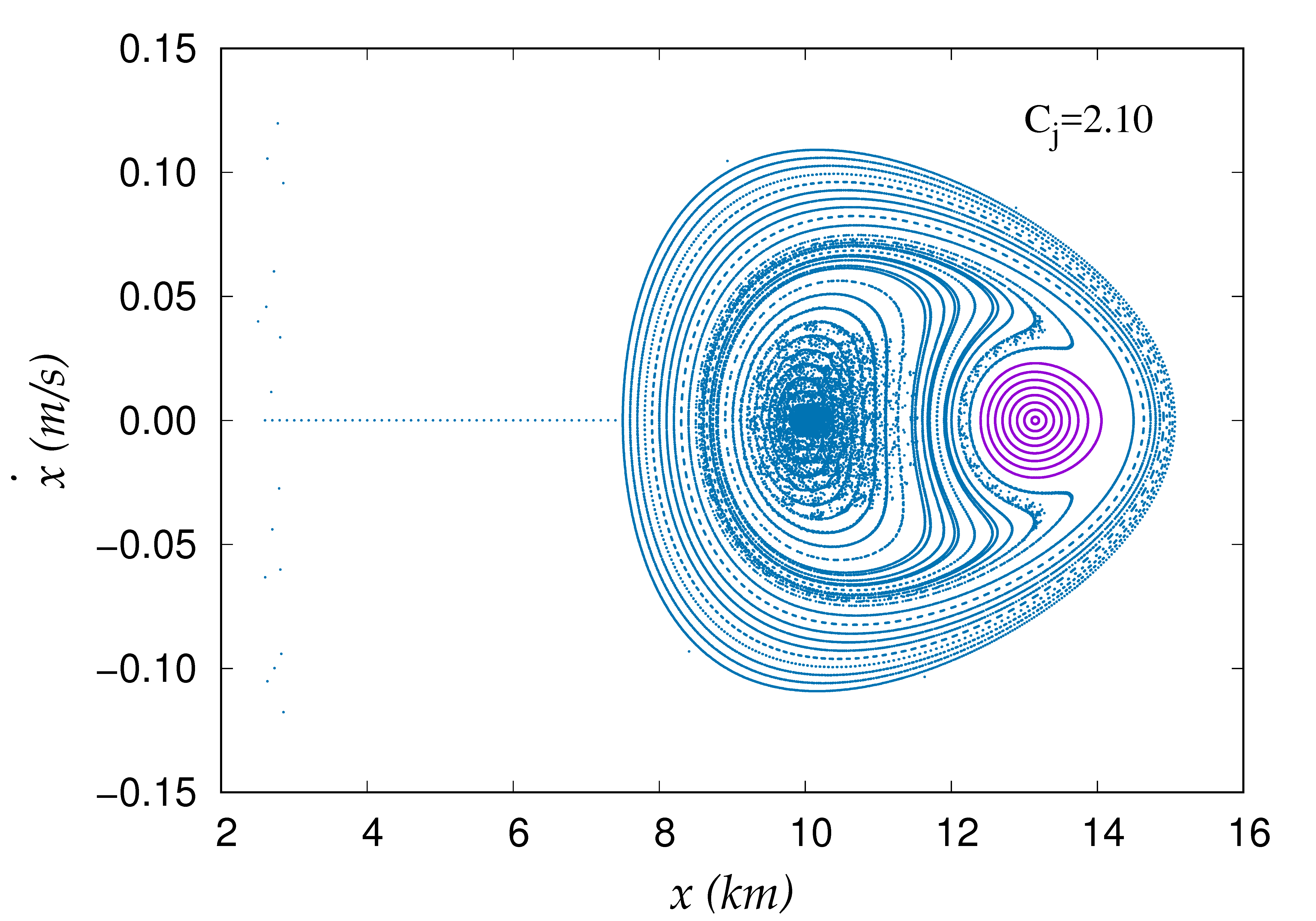}
	\includegraphics[width=14cm,height=7cm]{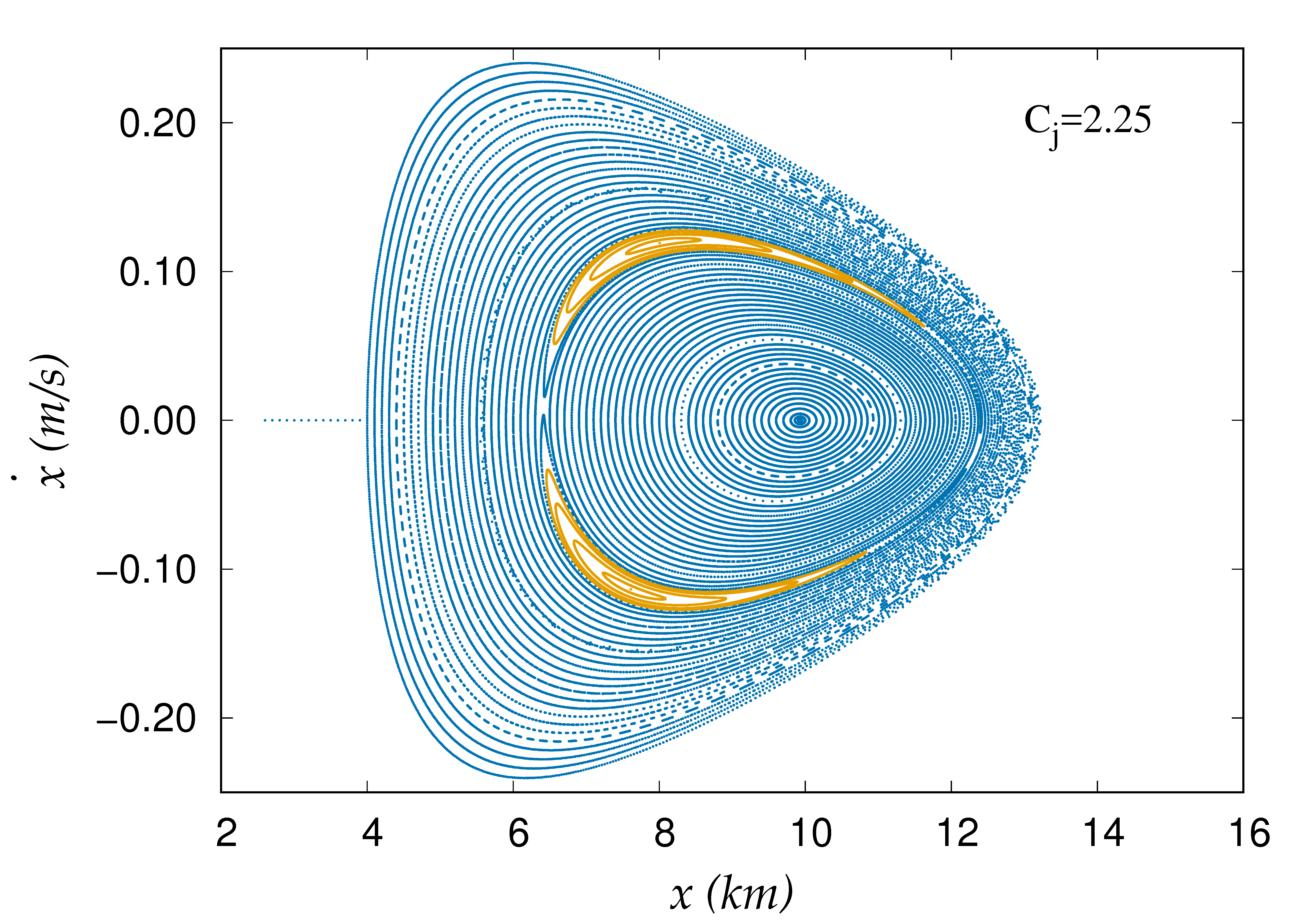}
    \caption{Poincar\'e surfaces of section for $C_{j}=1.20$, $2.10$ and $2.25$. There were set initial conditions with $x_{0}\geq 2.6$ $km$. Up to a thousand points per initial condition are generate for each plot. The curves associated to Family 1 are indicated in purple, Family 2 in orange and Family 3 in red.}
    \label{fig:4sp}
\end{figure*}

An illustrative sample of Poincar\'e surfaces of section is shown in Fig. \ref{fig:4sp}. Three Families of periodic orbits (projections of a \textit{central orbit}) were selected to be a representative sample of our results to be analyzed in detail. These structures are \textit{central orbits} and quasi-periodic orbits that librate around the \textit{central orbits}. The Families are presented in crescent order of complexity. The islands associated to Family 1 are indicated in purple, to Family 2 in orange, and to Family 3 in red.

There is a line of points with $\dot{x}=0$ in Fig. \ref{fig:4sp}, $C_{j}=2.10$ and $2.25$. These points indicate initial conditions whose trajectories had collided with the asteroid before the first complete orbital cycle in the rotating frame. The Poincar\'e surface of section for $C_{j}=1.20$ shows a line of points with $\dot{x}=0$ corresponding to unstable initial conditions whose trajectories have been ejected from the frame. The ejection condition is reached when the particle is further than  $~250$ $km$ from the center of the asteroid. The larger is the $C_{j}$ value, the smaller is the chaotic region on the Poincar\'e surface of section. 

In \citet{JiangSSP} two Poincar\'e surfaces of section are built in the potential of the asteroid 216 Kleopatra to show the chaotic behaviors of the orbits in large scale. This behavior is similar to the chaotic region seen in the present Poincar\'e surfaces of sections, despite of the irregular body and the dynamic region being different from the present case.

The zero-velocity surfaces can be useful to understand this behavior. These surfaces create forbidden regions when it is projected over the $xy$ plane. For $C_{j}=1.20$ the forbidden region is open(Fig. \ref{fig:open}). It allows large chaotic regions. On the other hand, for $C_{j}>2.003$ the forbidden region is closed(Fig. \ref{fig:closed}). It confines the movement,  increasing the stability regions. This behavior can be seen in more details in \citet{atlas1} and \citet{murraybook}. 

In this section, each selected Family will be studied. The effect of non-symmetry of the gravitational potential  on the structures from the Poincar\'e surface of section are analyzed as well.

These periodic orbits were found and followed through the Poincar\'e Surface of Section for different values of Jacob constant. Note that there are other approaches for doing that, like the continuation method as largely adopted in the literature \citep{YBJ,NI}.
%%%%%%%%%%%%%%%%%%%%%%%%%%%%%%%%%%%%%%%%%%%%%%%%%%%%%%%%%%%%%%%%%%%%%%%%%%%%%%%%%%%%%%%%%%%%%%%%%%%%
%%%%%%%%%%%%%%%%%%%%%%%%%%%%%%%%%%%%%%%%%%%%%%%%%%%%%%%%%%%%%%%%%%%%%%%%%%%%%%%%%%%%%%%%%%%%%%%%%%%%
%%%%%%%%%%%%%%%%%%%%%%%%%%%%%%%%%%%%%%%%%%%%%%%%%%%%%%%%%%%%%%%%%%%%%%%%%%%%%%%%%%%%%%%%%%%%%%%%%%%%
%%%%%%%%%%%%%%%%%%%%%%%%%%%%%%%%%%%%%%%%%%%%%%%%%%%%%%%%%%%%%%%%%%%%%%%%%%%%%%%%%%%%%%%%%%%%%%%%%%%%
\subsection{Family 1}

%%%%%%%%%%%%%%%%%%%%%%%%%%%%%%%%%%%%%%%%%%%%%%%%%%%%%%%%%%%%%%%%%%%%%%%%%%%%%%%%%%%%%%%%%%%%%%%%%%%%%%%%%%%%%%%%%%%%%%%%%%%%%%%%%%%%%%%%%%%%

\begin{figure}
\begin{center}
\subfloat{	\includegraphics[width=\columnwidth,trim = 10mm 0mm 10mm 0mm,clip]{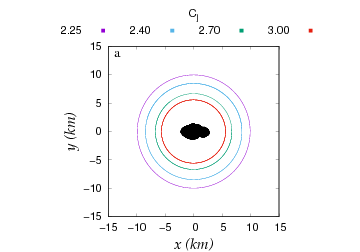}
\label{fig:f1-a-a}
}

\subfloat{	\includegraphics[width=\columnwidth]{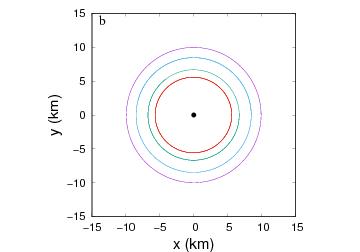}
\label{fig:f1-a-b}
}
\end{center}
   \caption{Sample of \textit{central orbits} of Family 1: a- in the rotating frame ($xy$); b- in the the inertial frame (xy). The orbits completed just one cycle in the rotating frame. The colors correspond to the indicated values of $C_{j}$.}
    \label{fig:f1-a}
\end{figure}

Family 1 is the most present one on the Poincar\'e surfaces of section for the studied values of $C_{j}$. Fig. \ref{fig:f1-a} shows a sample of  \textit{central orbits} of Family 1 projected on the rotating frame and the inertial frame. These orbits are almost circular in both, the rotating frame (Fig. \ref{fig:f1-a-a}) and the inertial frame (Fig. \ref{fig:f1-a-b}). The \textit{central orbits} are closer to the asteroid as the values of $C_{j}$ increases.

\begin{figure}
\subfloat{	\includegraphics[width=8 cm]{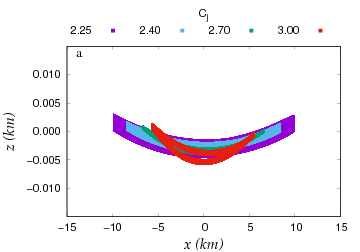}
   \label{fig:f1-b-a}
	}

\subfloat{	\includegraphics[width=8 cm]{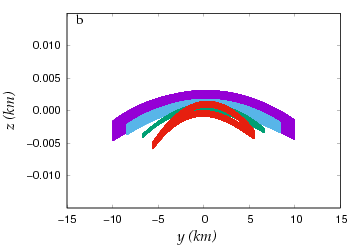}
   \label{fig:f1-b-b}
	}
   \caption{Sample of \textit{central orbits} of Family 1: a- projection of the trajectories in $xz$ plane; b- projection of the trajectories in $yz$ plane, both are presented in the rotating frame. The orbits completed many cycles in the rotating frame. The colors correspond to the indicated values of $C_{j}$.}
    \label{fig:f1-b}
\end{figure}

Fig. \ref{fig:f1-b} shows the projections of the \textit{central orbits} over the $xz$ and $yz$ planes. The variation amplitude in the $z$ axis is much smaller than the variation of the amplitude in the rotating frame. For higher values of $C_{j}$, the libration ranges of the \textit{central orbits} in the $z$ axis are decreasing.

\begin{figure}
	\includegraphics[width=8 cm]{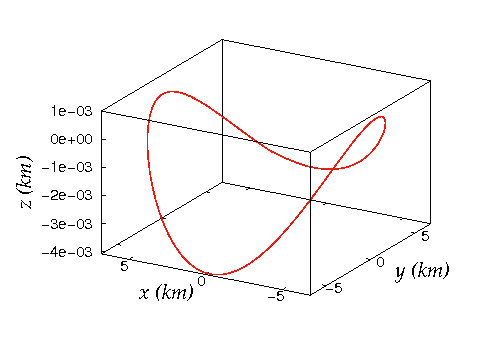}

   \caption{The \textit{central orbit} with $C_{j}=2.25$.  The shape is similar to the edges of a hyperbolic paraboloid. The trajectory completed many cycles in the rotating frame.}
    \label{fig:f1-hp}
\end{figure}

In the three-dimensional rotating space the \textit{central orbits} have the shape of the contour of a hyperbolic paraboloid surface in the rotating frame(Fig. \ref {fig:f1-hp}). 

\begin{figure}

\subfloat{	\includegraphics[width=9 cm]{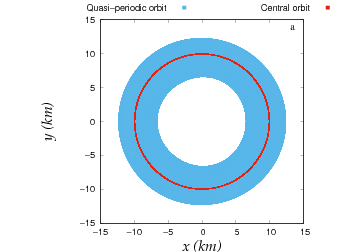}\\
\label{fig:f1-c-a}
}

\subfloat{	\includegraphics[width=9 cm]{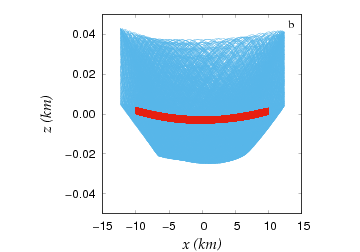}\\
\label{fig:f1-c-b}
}

\subfloat{	\includegraphics[width=9 cm]{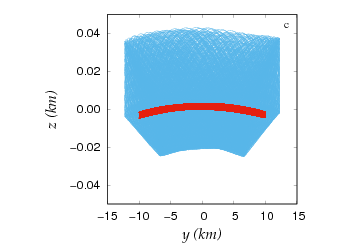}\\
\label{fig:f1-c-c}
}

    \caption{Projections of the \textit{central orbit}(red) and a large quasi-periodic orbit(blue) of Family 1. a- in the $xy$ plane; b- in the $xz$ plane; c- in the $yz$ plane. The Jacobi constant for these orbits is $2.25$.}
    \label{fig:f1-c}
\end{figure}

In order to illustrate the quasi-periodic orbits in Fig. \ref{fig:f1-c} is presented the libration of a large quasi-periodic orbit around a  \textit{central orbit} of the Family 1 with $C_{j}=2.25$. The libration conserve original structure of the \textit{central orbit}. Figs. \ref{fig:f1-c-b} and \ref{fig:f1-c-c} show the libration of a large quasi-periodic orbit in $xz$ and $yz$ plane, respectively. They are asymmetric in relation to the \textit{rotating plane} and the behavior is the same for the other families studied in this work. This characteristic will be analyzed in subsection \ref{3deffect}.  

\begin{figure}
	\includegraphics[width=\columnwidth]{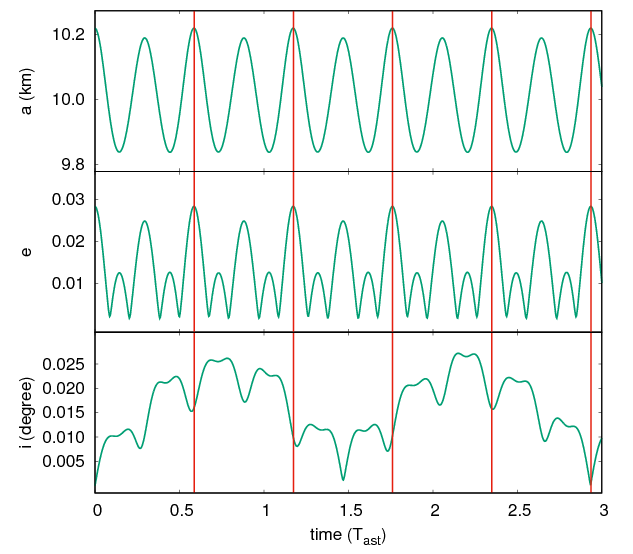}\\
    \caption{Temporal evolution of the semimajor-axis, eccentricity and inclination of Family 1 \textit{central orbit} with $C_{j}=2.25$. The red lines indicate a complete projected orbit period in the rotating frame.}
    \label{fig:f1-d}
\end{figure}

In Fig. \ref{fig:f1-d} the temporal evolution of the semimajor axis, eccentricity, and inclination of a \textit{central orbit} of Family 1 is presented. The semimajor axis and eccentricity have the same main frequency of variation(whose period is $\sim 0.586$ $T_{ast}$). The orbital inclination presents two main frequencies of periods $\sim 0.267T_{ast}$ and $\sim 1.469T_{ast}$.

\begin{figure}
	\includegraphics[width=\columnwidth]{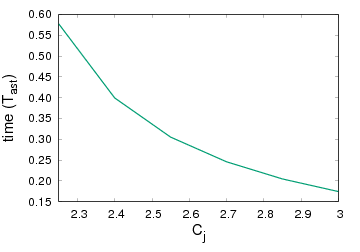}\\
    \caption{Evolution of the orbital period for Family 1 \textit{central orbits} when projected in the rotating frame as a function of the Jacobi constant.}
    \label{fig:f1-e}
\end{figure}

The periods of the \textit{central orbits} of Family 1 which are projected in the \textit{rotating plane} in function of Jacobi constant are shown in Fig. \ref{fig:f1-e}. The period decreases as the value of $C_{j}$ is increased. This period evolution is continuous between $0.17$ and $0.58$ $T_{ast}$ and is not connected to a resonance, even crossing values commensurable  with the asteroid period(Fig. \ref{fig:f1-e}). As the \textit{central orbits} are not in a given resonance and the eccentricity are very low, Family 1 is classified as of the first kind \citep{poincar}. Other classification adopted by \citet{Jiang2015cl} and \citet{Jiang2016cl} classifies periodic orbits using the position, the geometrical, and the topological characteristics.
%%%%%%%%%%%%%%%%%%%%%%%%%%%%%%%%%%%%%%%%%%%%%%%%%%%%%%%%%%%%%%%%%%%%%%%%%%%%%%%%%%%%%%%%%%%%%%%%%%%%
%%%%%%%%%%%%%%%%%%%%%%%%%%%%%%%%%%%%%%%%%%%%%%%%%%%%%%%%%%%%%%%%%%%%%%%%%%%%%%%%%%%%%%%%%%%%%%%%%%%%
%%%%%%%%%%%%%%%%%%%%%%%%%%%%%%%%%%%%%%%%%%%%%%%%%%%%%%%%%%%%%%%%%%%%%%%%%%%%%%%%%%%%%%%%%%%%%%%%%%%%
%%%%%%%%%%%%%%%%%%%%%%%%%%%%%%%%%%%%%%%%%%%%%%%%%%%%%%%%%%%%%%%%%%%%%%%%%%%%%%%%%%%%%%%%%%%%%%%%%%%%
\subsection{Family 2}

\begin{figure*}
	\includegraphics[width=14cm,height=7cm]{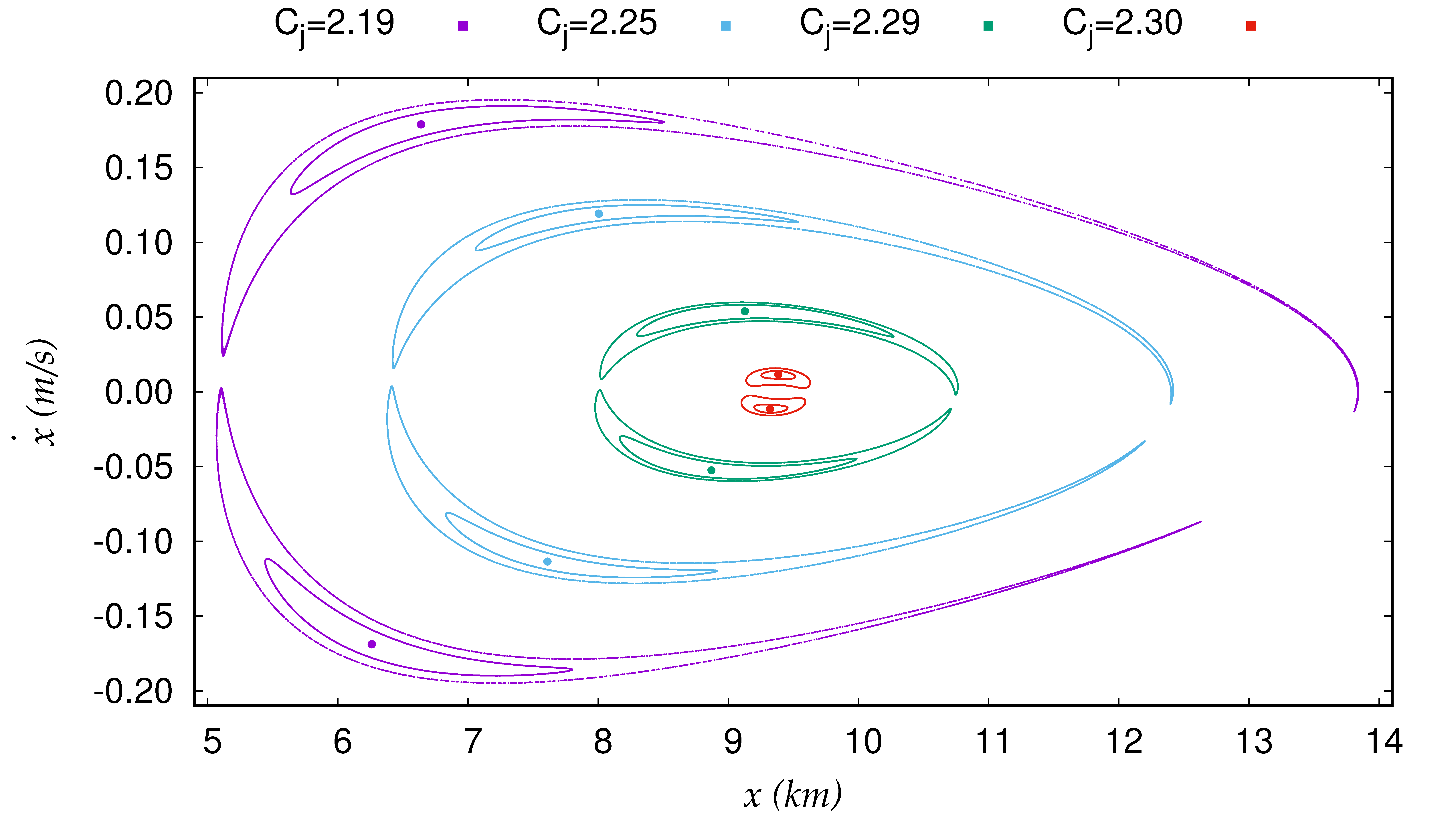}\\
    \caption{The evolution of Family 2 in Poincar\'e surfaces of section. A sample of the largest stability islands, a pair of intermediary islands and the dots inside that represent the \textit{central orbit} showing the structure for different values of $C_{j}$. The colors correspond to the indicated values of $C_{j}$.}
    \label{fig:f2-a}
\end{figure*}

Family 2 is identified by a pair of stable islands for values of $C_{j}$ from  $2.19$ to $2.30$. These islands are around two points in the Poincar\'e surfaces of section and these two points represents the \textit{central orbit} of Family 2. Fig. \ref{fig:f2-a} shows the evolution of this family using the largest pair of islands, a medium pair of islands, and a pair of dots from a sample of different values of $C_{j}$. For large values of $C_{j}$ the islands come closer to each other and the amplitude of the variation(size of the islands) decreases. This behavior is reflected in the orbital eccentricity of the trajectories. In Fig. \ref{fig:f2-a} an asymmetry is identified. Such asymmetry is created by the non-symmetry of the gravitational potential. This behavior persists even when the effect of the third dimension is neglected. 

For values of $C_{j}$ lower than $2.19$ and higher than $2.30$ this Family does not exist. For  $C_{j}<2.19$, the structure should be in a region whose trajectories collide with the asteroid. 

\begin{figure}
	\includegraphics[width=\columnwidth]{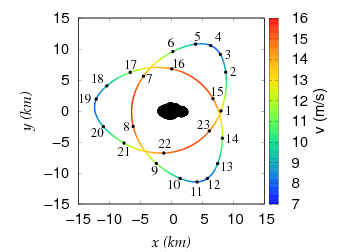}\\
    \caption{The \textit{central orbit} of Family 2 in rotating frame for Jacobi constant $2.25$. The colors indicate the velocity in the inertial frame. The orbit is divided in 23 parts with equal time steps and the points are numbered to show the trajectory sequence.}
    \label{fig:f2-b}
\end{figure}

In Fig. \ref{fig:f2-b} the trajectory of the  \textit{central orbit} with $C_{j}=2.25$ in the rotating frame is presented. The black points are equally spaced in time and the colors indicate the modulus of the velocity in the inertial frame. The higher velocities coincide with the moment in the trajectories whose particle is closer to the asteroid, characterizing the three pericenters. On the other hand, regions with lower velocity and away from the asteroid  characterize the three apocenters.

\begin{figure}

\subfloat{	\includegraphics[width=\columnwidth]{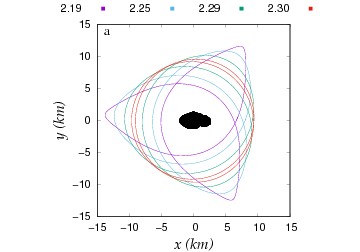}
\label{fig:f2-c-a}
}

\subfloat{	\includegraphics[width=\columnwidth]{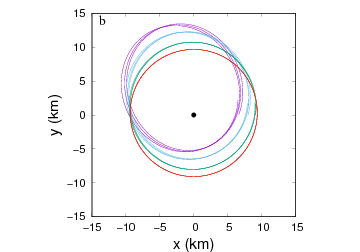}\\
\label{fig:f2-c-b}
} 
  \caption{\textit{Central orbits} from Family 2: a- rotating frame, $xy$ plane; b- inertial frame, xy plane. The orbits completed just one cycle in the rotating frame. The colors correspond to the indicated values of $C_{j}$.}
    \label{fig:f2-c}
\end{figure}

In Fig. \ref{fig:f2-c} the evolution of a representative sample of \textit{central orbits} in function of $C_{j}$ is presented. As far as the value of $C_{j}$ is increased, the eccentricity decreases.

\begin{figure}
\subfloat{	\includegraphics[width=9 cm]{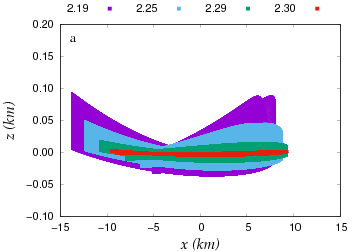}
 \label{fig:f2-d-a}
}

\subfloat{	\includegraphics[width=9 cm]{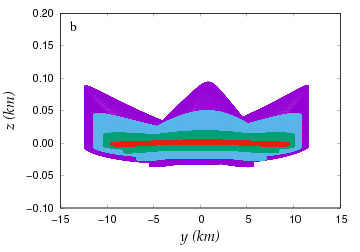}\\
 \label{fig:f2-d-b}
}

   \caption{\textit{Central orbits} from Family 2: a- projection of the trajectories in $xz$ plane; b- projection of the trajectories in $yz$ plane, both are presented in the rotating frame. The orbits completed many cycles at rotating frame. The colors correspond to different values of $C_{j}$.}
    \label{fig:f2-d}
\end{figure}

In Fig. \ref{fig:f2-d} the projections of \textit{central orbits} in $xz$ and $yz$ planes in the rotating frame are presented. The amplitude variation in the third dimension is much smaller than the amplitude variation in the other two dimensions. Furthermore, all variations decrease when the value of $C_{j}$ is increased.

\begin{figure}
	\includegraphics[width=\columnwidth]{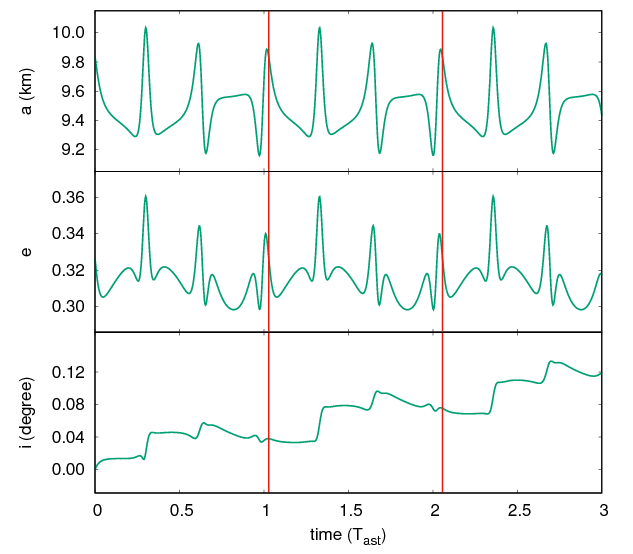}\\
    \caption{The evolution of the semimajor axis, eccentricity and inclination of Family 2 \textit{central orbit} with $C_{j}=2.25$. The red lines correspond to a complete orbit period in the rotating frame.}
    \label{fig:f2-e}
\end{figure}

\begin{figure}
	\includegraphics[width=\columnwidth]{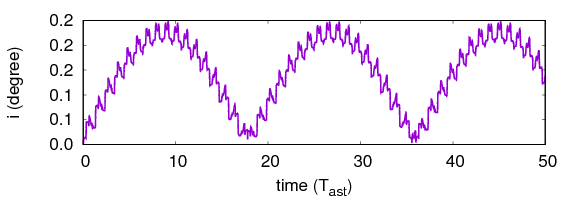}\\
    \caption{Evolution of the orbital inclination of Family 2 \textit{central orbit} with $C_{j}=2.25$.}
    \label{fig:f2-f}
\end{figure}

The evolution of the semimajor-axis, the eccentricity, and the inclination of the \textit{central orbit} of Family 2 with  $C_{j}=2.25$ are presented in Fig. \ref{fig:f2-e}. It is possible to identify the same main frequency (whose period is  $\sim 1.028$ $T_{ast}$) in both the semimajor-axis and the eccentricity. The inclination has two main frequencies. The frequency with the largest period($\sim 17.79$ $T_{ast}$) is better seen in Fig. \ref{fig:f2-f}.

\begin{figure}
	\includegraphics[width=\columnwidth]{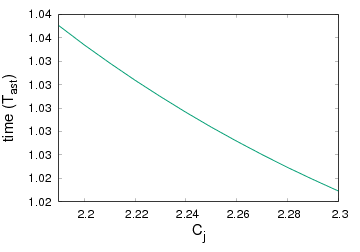}\\
    \caption{Evolution of the orbital period for Family 2 \textit{central orbits} when projected in the rotating frame as a function of the Jacobi constant.}
    \label{fig:f2-g}
\end{figure}

The projection of the \textit{central orbit} is a periodic orbit. The period of this periodic orbit for different values of $C_{j}$ is computed, then a curve of the evolution of that period is built(Fig. \ref{fig:f2-g}). The periods computed are close to the rotational period of the asteroid. In Figs. \ref{fig:f2-b} and \ref{fig:f2-c}, each complete cycle in the rotating frame corresponds approximately to three complete cycles in the inertial frame. Therefore, Family 2 is associated to the 3:1 resonance(particle orbit : asteroid rotation).

\begin{figure*}

\subfloat{	\includegraphics[width=\columnwidth,trim = 5mm 0mm 5mm 0mm,clip]{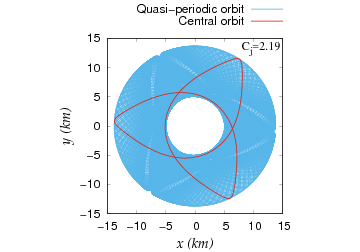} }
\subfloat{	\includegraphics[width=\columnwidth]{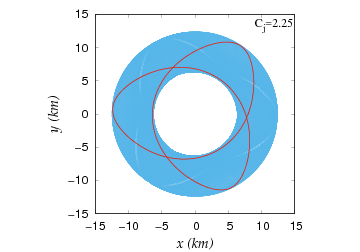} } 

\subfloat{	\includegraphics[width=\columnwidth]{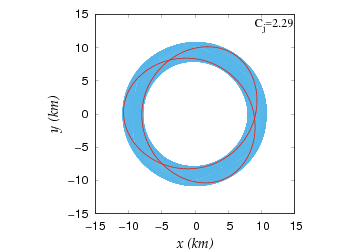} }
\subfloat{	\includegraphics[width=\columnwidth]{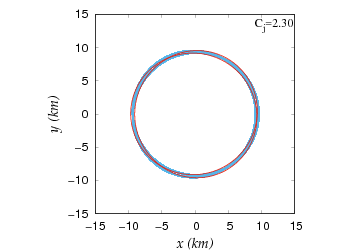} }

    \caption{The largest quasi-periodic orbit (blue) and the \textit{central orbit} (red) of Family 2 in the rotating frame for different values of $C_{j}$.}
   \label{fig:f2-h}
\end{figure*}

The largest quasi-periodic orbit is used to define the width of the region where the structure of the trajectories are similar to the \textit{central orbit}(figura \ref{fig:f2-h}). Higher values of $C_{j}$ present \textit{central orbits} with lower eccentricity and smaller libration amplitude.

\citet{Sh98}(see their Fig. $13b$) found a periodic orbit similar to a \textit{central orbit} of Family 2. When using the Poincar\'e surface of section we are showing that it is possible to find not only the whole  Family but also the width of the libration regions.
%%%%%%%%%%%%%%%%%%%%%%%%%%%%%%%%%%%%%%%%%%%%%%%%%%%%%%%%%%%%%%%%%%%%%%%%%%%%%%%%%%%%%%%%%%%%%%%%%%%%
%%%%%%%%%%%%%%%%%%%%%%%%%%%%%%%%%%%%%%%%%%%%%%%%%%%%%%%%%%%%%%%%%%%%%%%%%%%%%%%%%%%%%%%%%%%%%%%%%%%%
%%%%%%%%%%%%%%%%%%%%%%%%%%%%%%%%%%%%%%%%%%%%%%%%%%%%%%%%%%%%%%%%%%%%%%%%%%%%%%%%%%%%%%%%%%%%%%%%%%%%
%%%%%%%%%%%%%%%%%%%%%%%%%%%%%%%%%%%%%%%%%%%%%%%%%%%%%%%%%%%%%%%%%%%%%%%%%%%%%%%%%%%%%%%%%%%%%%%%%%%%
\subsection{Family 3}

\begin{figure*}
	\includegraphics[width=14cm,height=7cm]{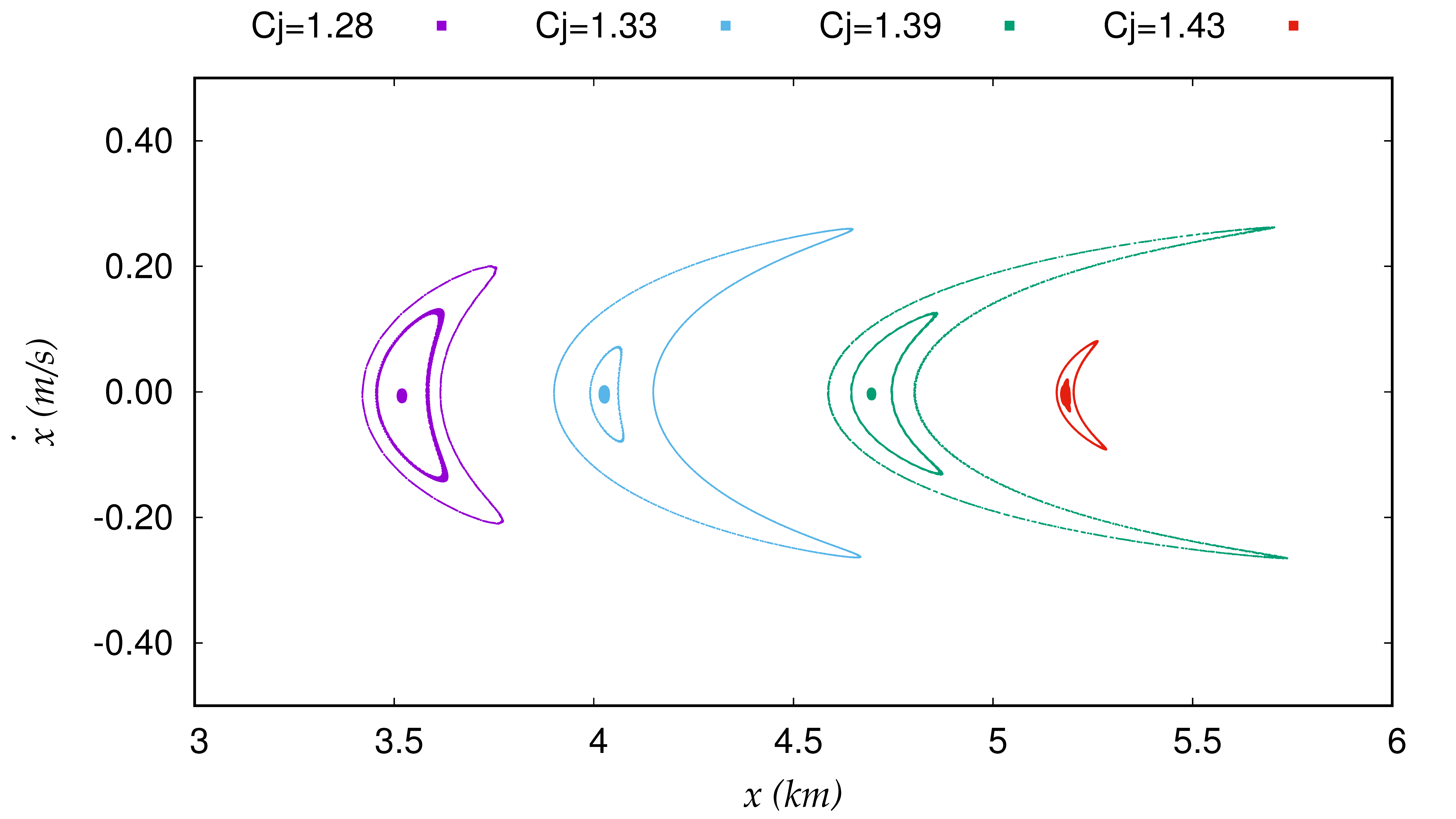}\\
    \caption{The evolution of Family 3 in Poincar\'e surfaces of section. A sample of the largest stability islands, intermediary islands and the dots that represent the \textit{central orbit} showing the structure for different values of $C_{j}$. The colors correspond to the indicated values of $C_{j}$.}
    \label{fig:f3-a}
\end{figure*}

Family 3 is in the Poincar\'e surface of section with lower values of $C_{j}$ among the values studied here. In Fig. \ref{fig:f3-a} the evolution of the family's structure on the Poincar\'e surfaces of section is presented with a sample of the largest island, a medium island and the point that corresponds to the \textit{central orbit}. These structures with different values of $C_{j}$ were chosen in order to be a representative sample. When the value of $C{j}$ is increased, the width of the island becomes larger but the island for the highest value of $C{j}$(in red) breaks this behavior. This phenomena is due to the chaotic region where this island is placed. The chaos destroys the islands that are larger than the island in Fig. \ref{fig:f3-a} with $C{j}=1.43$. 

\begin{figure}
	\includegraphics[width=\columnwidth]{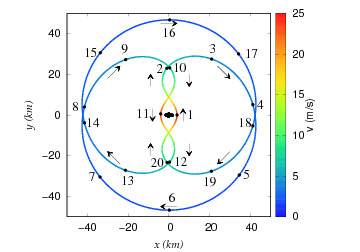}\\
    \caption{The \textit{central orbit} from the family 3 in rotating frame for Jacobi constant $1.33$. The colors indicate the velocity in the inertial frame. The orbit is divided in 20 parts with equal time steps and the points are numbered to show the trajectory sequence. To assist in the understanding of the trajectory, arrows are showing the direction of the orbit.}
    \label{fig:f3-b}
\end{figure}

\begin{figure}
\subfloat{	\includegraphics[width=\columnwidth]{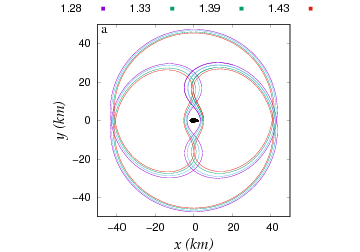}
\label{fig:f3-c-a}
}

\subfloat{	\includegraphics[width=\columnwidth]{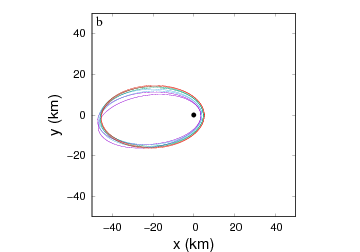}
\label{fig:f3-c-b}
}
   \caption{\textit{Central orbits} from Family 3: a- rotating frame, $xy$ plane; b- inertial frame, xy plane. The orbits completed just one cycle in the rotating frame. The colors correspond to the indicated values of $C_{j}$.}
    \label{fig:f3-c}
\end{figure}

Fig. \ref{fig:f3-b} shows the trajectory of the  \textit{central orbit} of Family 3 with $C_{j}=1.33$. The arrows give a sequential direction of the trajectory and the points are plotted with equal time step. There are two regions with higher velocity and closer to the asteroid, indicating the pericenters of the orbit. The opposite situation, i.e. lower velocities and farther from the asteroid, indicates the apocenters. A difference between Family 3 and the other families is that there are moments whose trajectories become retrograde in the rotating frame. It is due to the high orbital eccentricity. The variation of the orbital velocity is wide enough to be faster than the rotating velocity of the asteroid when it is close to the pericenter and slower when it is close to the apocenter. Fig. \ref{fig:f3-c} shows a sample of the \textit{central orbits} in the rotating frame and in the inertial frame. It is possible to note the existence of the two pericenters and the two apocenters.    

\begin{figure}
\subfloat{	\includegraphics[width=9 cm]{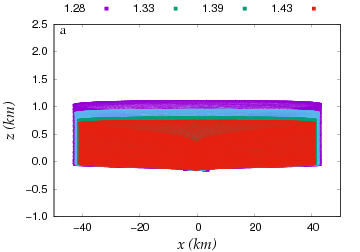}
\label{fig:f3-d-a}
}

\subfloat{	\includegraphics[width=9 cm]{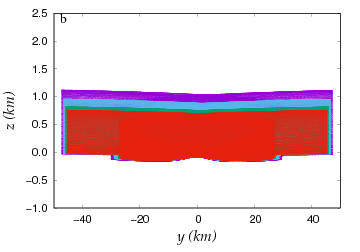}
\label{fig:f3-d-b}
}
   \caption{The \textit{central orbits} from Family 3: a- projection of the trajectories in $xz$ plane; b- projection of the trajectories in $yz$ plane, both are presented in the rotating frame. The orbits completed many cycles at the rotating frame. The colors correspond to different values of $C_{j}$.}
    \label{fig:f3-d}
\end{figure}

As in the other families, the amplitude of the variation in the $z$ axis is much smaller than the variation in the other axes. This amplitude decreases when the value of $C_{j}$ is increased(Fig. \ref{fig:f3-d}). The variation of the structure and of the amplitude of the trajectory in $xz$ and $yz$ planes are the smallest among the studied families. It is due to the fact that Family 3 is restricted to a small range of values of $C_{j}$.

\begin{figure}
	\includegraphics[width=\columnwidth]{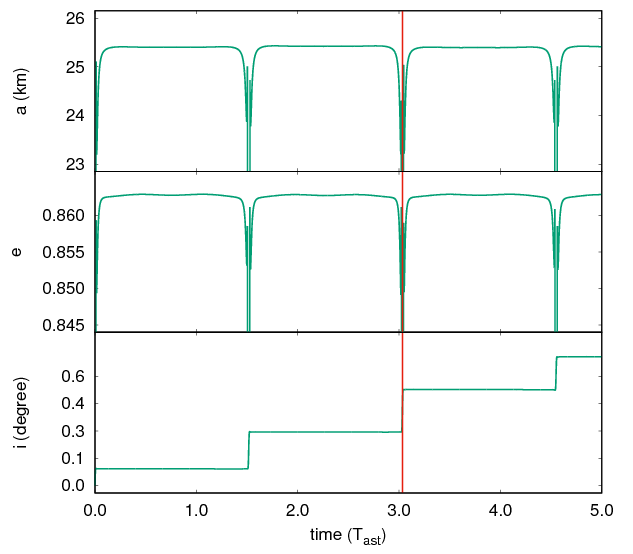}\\
    \caption{The evolution of the semimajor axis, eccentricity and inclination of Family 3 \textit{central orbit} with $C_{j}=1.28$. The red line corresponds to a complete projected orbit period in the rotating frame.}
    \label{fig:f3-e}
\end{figure}

\begin{figure}
	\includegraphics[width=\columnwidth]{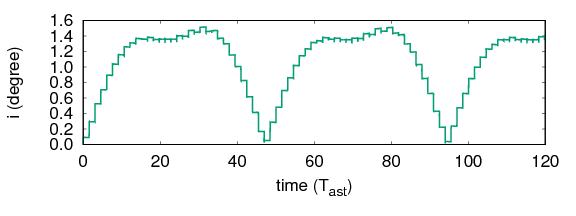}\\
    \caption{Evolution of the orbital inclination of Family 3 \textit{central orbit} with $C_{j}=1.28$.}
    \label{fig:f3-f}
\end{figure}

Fig. \ref{fig:f3-e} shows the evolution of semimajor-axis, eccentricity, and inclination of the \textit{central orbit} with $C_{j}=1.28$. The period of the main frequency of the semimajor-axis and of the eccentricity is $\sim 1.52$ $T_{ast}$ and the main frequency with the higher period from the inclination can be seen in Fig. \ref{fig:f3-f}, which is $\sim 47.07$ $T_{ast}$. 

\begin{figure}
	\includegraphics[width=\columnwidth]{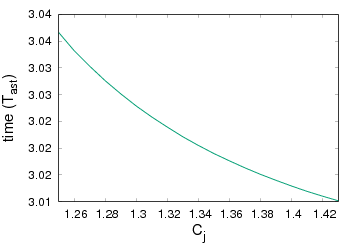}\\
    \caption{Evolution of the orbital period for Family 3 \textit{central orbits} when projected in the rotating frame as a function of the Jacobi constant.}
    \label{fig:f3-g}
\end{figure}

The period variation of the \textit{central orbits} for different values of $C_{j}$ is shown in Fig. \ref{fig:f3-g}, where the central orbit period is close to three rotational periods of the asteroid. Given that, a cycle in the rotating frame ($\sim 3 T_{ast}$) corresponds to two cycles in the inertial frame(Figs. \ref{fig:f3-b} e \ref{fig:f3-c}). It indicates a connection between Family 3 and the resonance 2:3(particle orbit : asteroid rotation).
 
\begin{figure*}
\subfloat{	\includegraphics[width=\columnwidth,trim = 5mm 0mm 5mm 0mm,clip]{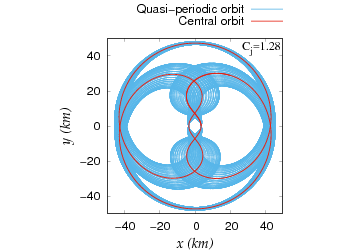}}
\subfloat{	\includegraphics[width=\columnwidth]{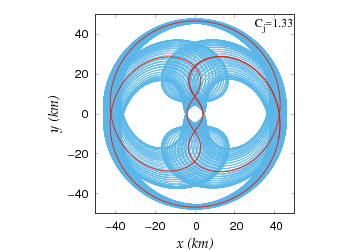}}

\subfloat{	\includegraphics[width=\columnwidth]{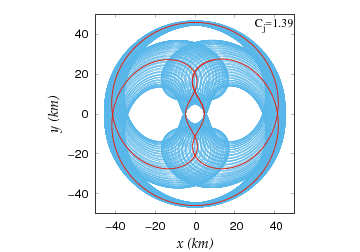}}
\subfloat{	\includegraphics[width=\columnwidth]{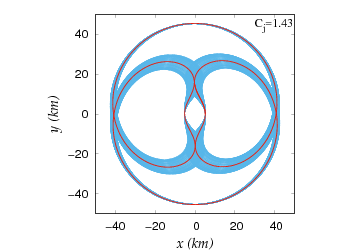}}
    \caption{The largest quasi-periodic orbit (blue) and the \textit{central orbit} (red) of Family 3 in the rotating frame for different values of $C_{j}$.}
    \label{fig:f3-h}
\end{figure*}

A sample of the largest quasi-periodic orbits around the \textit{central orbits} for different values of $C_{j}$ is presented in Fig. \ref{fig:f3-h}. This largest quasi-periodic orbits are the limit in which the orbits preserve the structure of the \textit{central orbit}. 
%%%%%%%%%%%%%%%%%%%%%%%%%%%%%%%%%%%%%%%%%%%%%%%%%%%%%%%%%%%%%%%%%%%%%%%%%%%%%%%%%%%%%%%%%%%%%%%%%%%%
%%%%%%%%%%%%%%%%%%%%%%%%%%%%%%%%%%%%%%%%%%%%%%%%%%%%%%%%%%%%%%%%%%%%%%%%%%%%%%%%%%%%%%%%%%%%%%%%%%%%
%%%%%%%%%%%%%%%%%%%%%%%%%%%%%%%%%%%%%%%%%%%%%%%%%%%%%%%%%%%%%%%%%%%%%%%%%%%%%%%%%%%%%%%%%%%%%%%%%%%%
%%%%%%%%%%%%%%%%%%%%%%%%%%%%%%%%%%%%%%%%%%%%%%%%%%%%%%%%%%%%%%%%%%%%%%%%%%%%%%%%%%%%%%%%%%%%%%%%%%%%
\subsection{The 3-D effect}
\label{3deffect}

Since we are dealing with a three-dimensional system, the analysis of the Poincar\'e surfaces of section alone is not enough to fully understand the dynamics. In the current section we introduce an analysis of the Poincar\'e surfaces of section combined with the amplitude of variation of the $z$ component of the trajectories. We generated plots that shown, for each initial value of $x$ ($x_{0}$), the upper and lower limits of the $z$ component among the points of Poincar\'e surfaces of section.

\begin{figure}
\begin{center}
\subfloat{	\includegraphics[width=\columnwidth]{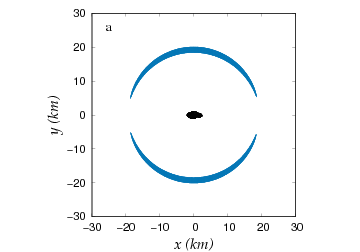}
\label{fig:open}
}

\subfloat{	\includegraphics[width=\columnwidth]{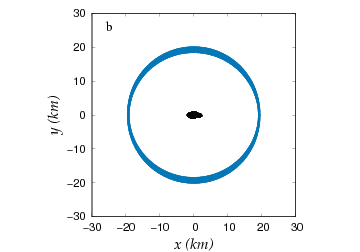}
\label{fig:closed}
}
\end{center}
   \caption{Illustrative cases of zero-velocity curves indicating  the forbidden region (in blue). To the case a- $C_{j}=2.003$ and the case b- $C_{j}=2.004$.}
    \label{fig:cv0}
\end{figure}

\begin{figure*}

\includegraphics[width=17cm,trim = 5mm 40mm 7mm 0mm,clip]{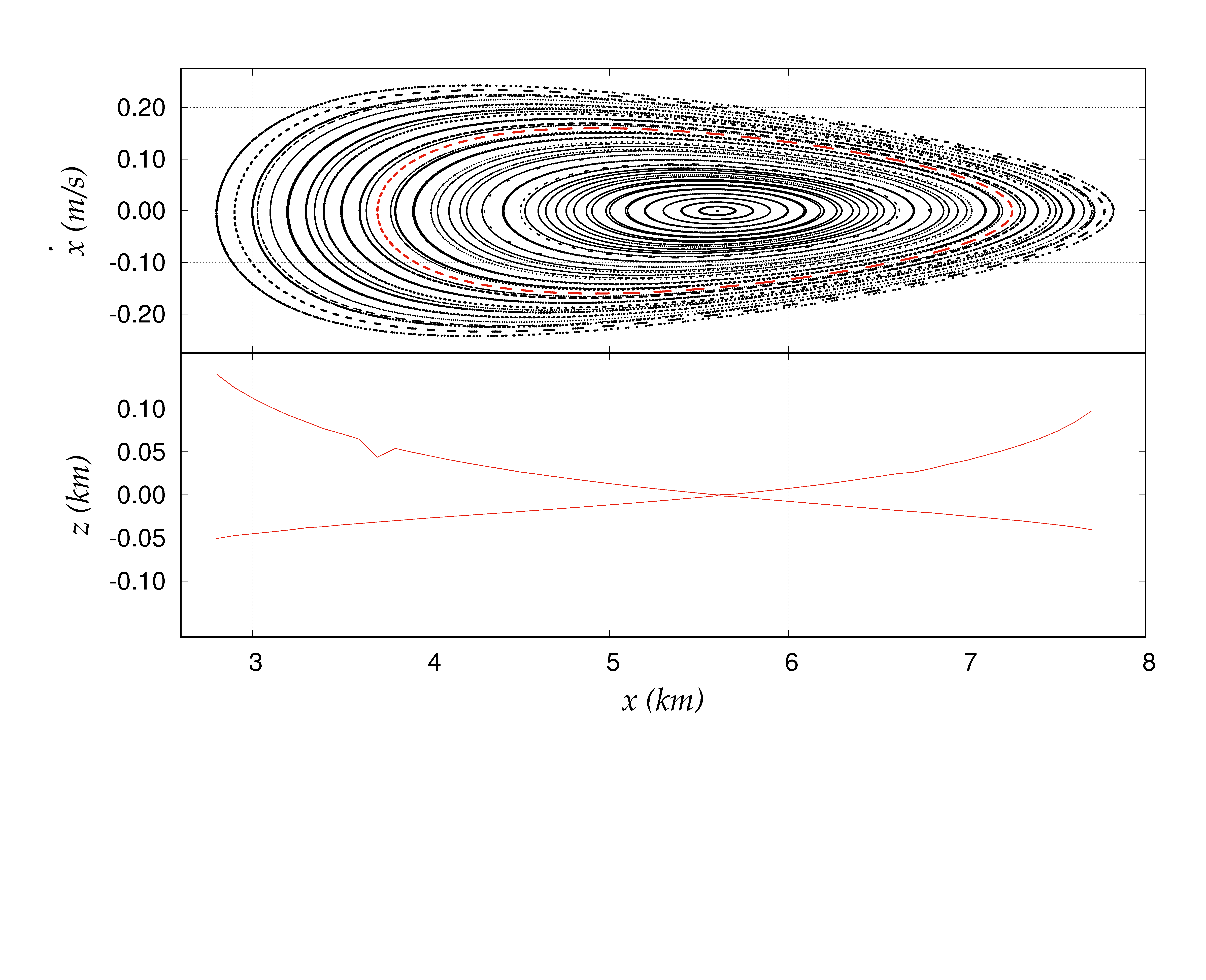}

\caption{Poincar\'e surface of section for $C_{j}=3.00$ and the limits of the variation in the $z$ axis for each initial condition. A cluster of stability islands for just one initial condition is indicated in red in the surface of section. }
    \label{fig:3d-a}
\end{figure*}

Following are presented a representative sample of the Poincar\'e surfaces of section and their limits of variation in the third dimension. Fig. \ref{fig:3d-a} shows the Poincar\'e surface of section for $C_{j}=3.00$(top) and the limits of variation in the third dimension for each initial condition in the $x$ axis(bottom). For this value of the Jacobi constant, the forbidden region is closed (Fig. \ref{fig:closed}). The initial conditions $x_{0}\geq 9.2$ $km$ are not integrated because they are inside the forbidden region. On the other hand, the initial conditions with $8.0 \leq x_{0}<9.2$ $km$ collide with the asteroid or are ejected. The limits of the variation in the third dimension are much lower than the amplitude of the trajectory in the $xy$ plane. When the initial condition is set closer to the  \textit{central orbit} of Family 1, the limit of variation in $z$ is closer to zero. In $x_{0}=3.7$ $km$ there is a break in the smoothness of the amplitude variation curve in the third dimension. In the Poincar\'e surface of section this initial condition (indicated in red) does not reflect any distinct behavior of its neighborhood, but when we integrate the trajectory for a longer time, it became visibly chaotic.

\begin{figure*}

\includegraphics[width=17cm,trim = 5mm 40mm 7mm 0mm,clip]{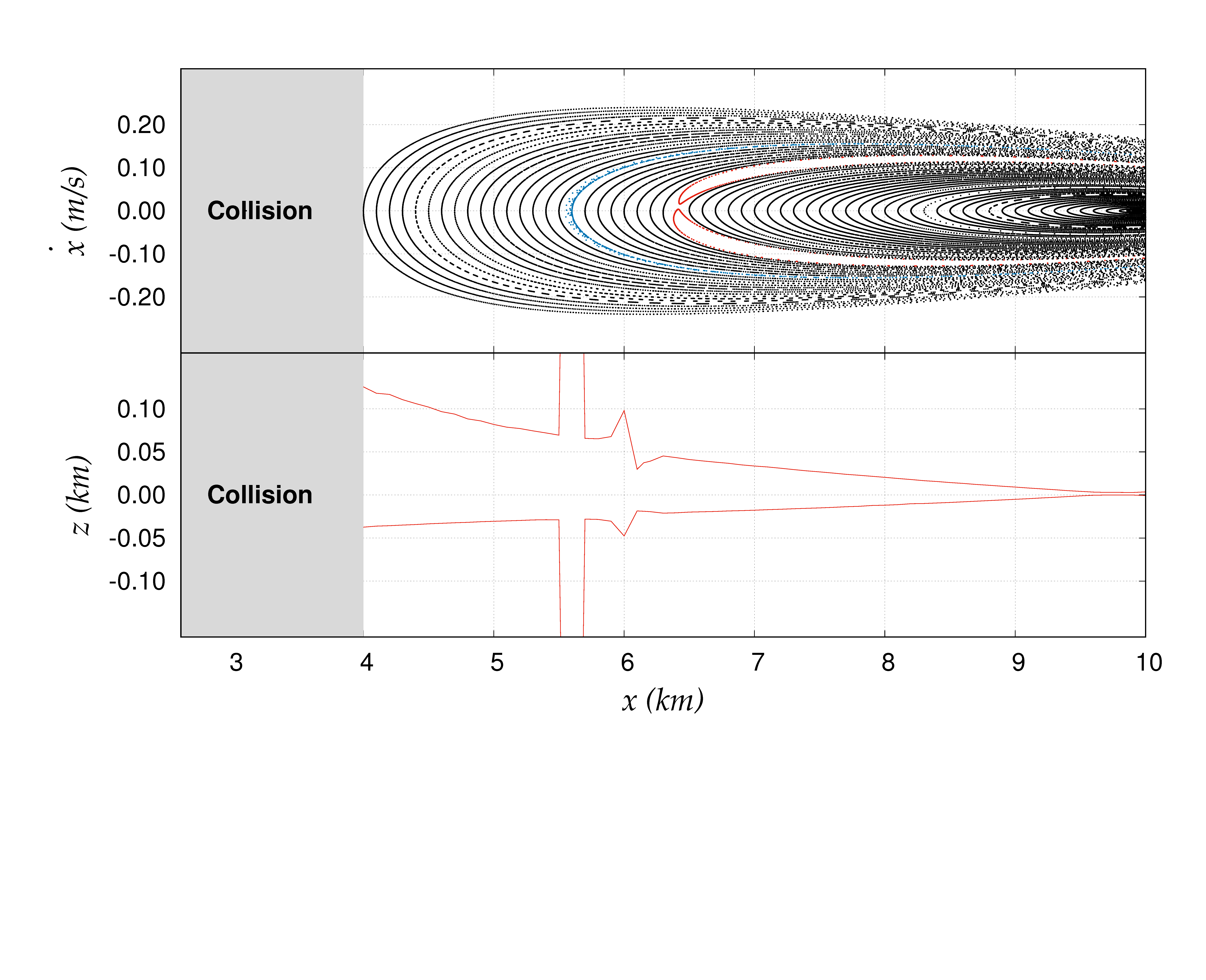}

\caption{Poincar\'e surface of section for $C_{j}=2.25$ and the limits of the variation in the $z$ axis for each initial condition. A chaotic trajectory is indicated in blue in the surface of section and a pair of islands in red.}
    \label{fig:3d-b}
\end{figure*}

Fig. \ref{fig:3d-b} shows the Poincar\'e surface of section and the limits of the variation in the third dimension for $C_{j}=2.25$. For this value of $C_{j}$, the forbidden region is still closed(Fig. \ref{fig:closed}), but smaller than the regions for higher values of $C_{j}$. In Fig. \ref{fig:3d-b} the region whose initial conditions have the trajectories interrupted by the collision of the orbiting particle with the asteroid are marked in gray. Close to the  \textit{central orbit} the limits in the third dimension tend to zero, as in the previous case. Some different structures are found there. In $x_{0}=5.6$ $km$ the points indicated in blue in the Poincar\'e surface of section correspond to the upper limit $\sim 1.18$ $km$ and the lower limit $\sim -1.10$ $km$ in the third dimension. These limits are outside the graph because they are much higher than the other values. Just a discontinuity is seen in the graph, which corresponds to a chaotic trajectory in the Poincar\'e surface of section. Therefore, the chaotic trajectory leads to a significant orbital variation in the third dimension, and vice versa. In this Poincar\'e surface of section with the value $C_{j}=2.25$ there is a \textit{central orbit} of the Family 2. In  $x_{0}=6.4$ is a quasi-periodic orbit that is librating around the \textit{central orbit} of Family 2, indicated in red. Note that the structure does not have a variation in the behavior of the limit in the third dimension, since it is not chaotic.

\begin{figure*}

\includegraphics[width=17cm,trim = 5mm 40mm 7mm 0mm,clip]{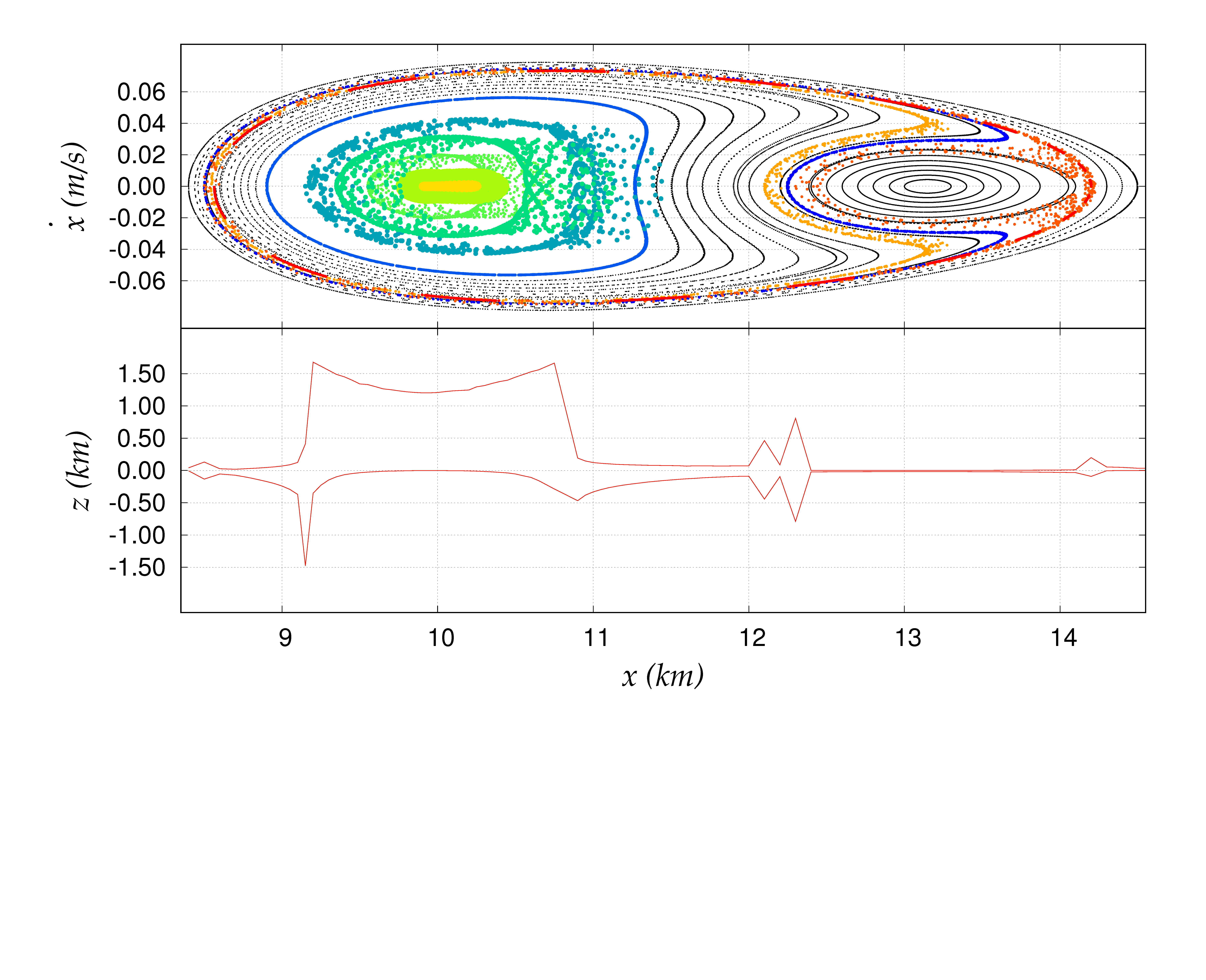}

\caption{Poincar\'e surface of section for $C_{j}=2.10$ and the limits of the variation in the $z$ axis for each initial condition. The chaotic structures are indicated in different colors for each initial condition. }
    \label{fig:3d-c}
\end{figure*}

Fig. \ref{fig:3d-c} shows the Poincar\'e surface of section for $C_{j}=2.10$ and the respective limits of the variation in the third dimension. For this value of $C_{j}$, the forbidden region is still closed. The Poincar\'e surface of section presents stable and unstable regions in equal proportion. There are structures with  two \textit{central orbits}. Between $x_{0}=12.4$ and $14.1$ $km$, the limits of the $z$ variation are very close to zero compared to the others in the graph and such behavior is similar to one in Fig. \ref{fig:3d-a}. This region corresponds to a stable region of quasi-periodic orbits around the \textit{central orbit} of Family 1. Other stable regions can be found  by the evolution of the limits in the third dimension and correspond to quasi-periodic orbits. The quasi-periodic orbits are indicated in black in the Poincar\'e surface of section. Peaks in  $x_{0}=8.5$ and $14.2$ $km$ correspond to chaotic trajectories in the Poincar\'e surface of section. The results for $x_{0}=8.5$ $km$ are indicated in dark blue and the results for $x_{0}=14.2$ $km$ are indicated in red. From $x_{0}=12.0$ to $12.4$ $km$ there is an unstable region(orange) confined between two stable regions. And between $x_{0}=9.1$ and $10.9$ $km$ there is an unstable region with higher limits of variations in the third dimension. In this region each initial condition is indicated in different colors(yellow, green, and blue). There is a chaotic structure confined within this region. This structure presents a continuous behavior in the Poincar\'e surface of section, and in the limits of variation in the third dimension. In the Poincar\'e surface of section at least one trajectory crosses from unstable region to stable region. This phenomena is just apparent. Actually, due to the third dimension. In the Poincar\'e surface of section of two-dimensional systems this does not occur. The two-dimensional case accepts just one solution of $x$ and $\dot{x}$ by the Picard$-$Lindel\"of theorem. Therefore, there is not a crossing between the two solutions for different initial conditions in a two-dimensional case. Here we study a three-dimensional case projected in a two-dimensional phase space. There is just one solution for $x$, $z$, $\dot{x}$ and $\dot{z}$, but infinite number of solutions for $x$ and $\dot{x}$ with different values of $z$ and $\dot{z}$. 

\begin{figure*}

\includegraphics[width=17cm,trim = 5mm 40mm 7mm 0mm,clip]{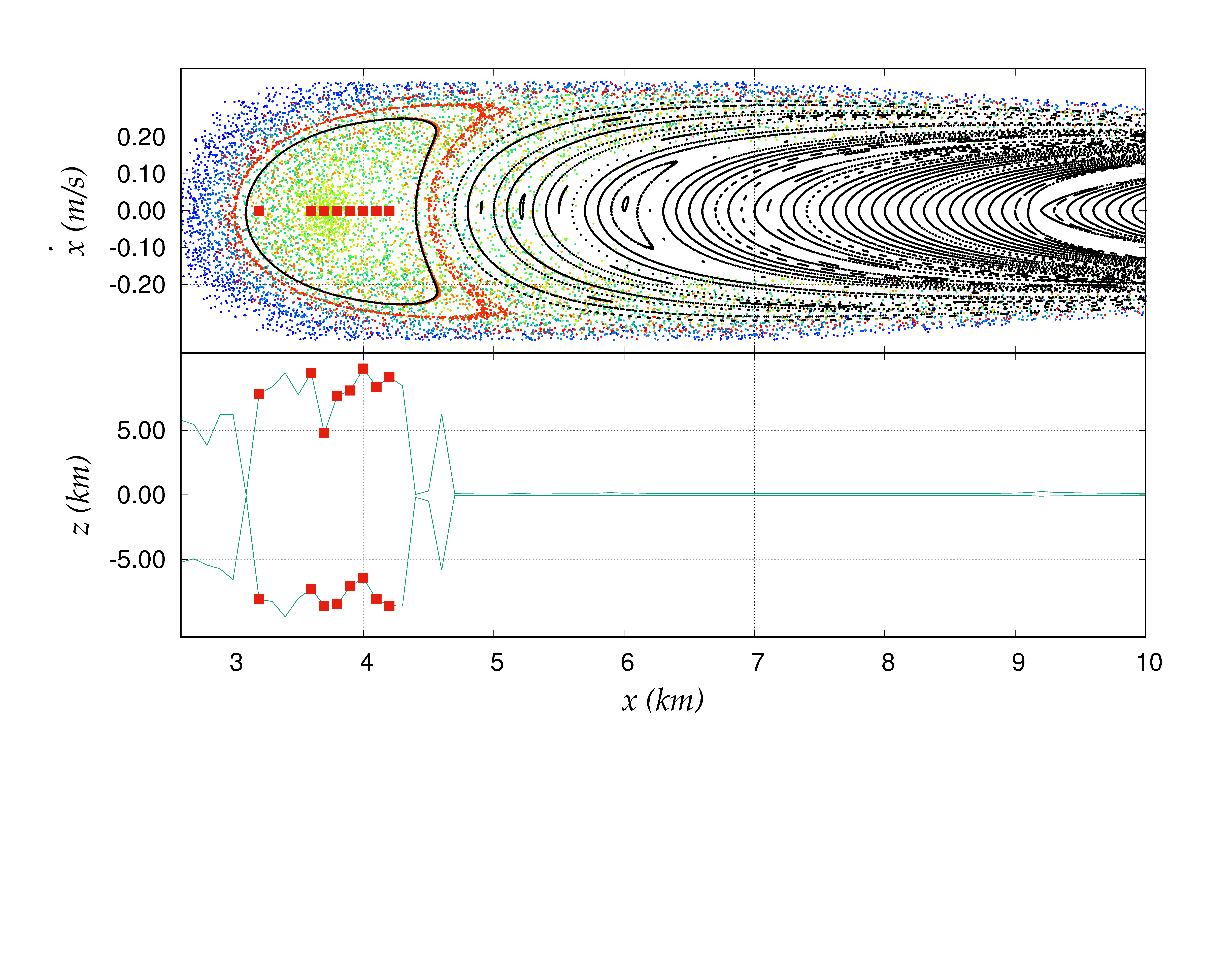}

\caption{Poincar\'e surface of section for $C_{j}=1.80$ and the limits of the variation in the $z$ axis for each initial condition. The chaotic structures are indicated in different colors for each initial condition. Trajectories that collided with the asteroid are indicated by the red squares. }
    \label{fig:3d-d}
\end{figure*}

In Fig. \ref{fig:3d-d} the Poincar\'e surface of section for $C_{j}=1.80$ and the respective limits of the movement in third dimension are presented. For this value of $C_{j}$, the forbidden region is opened (Fig. \ref{fig:closed}). The stable trajectories are indicated in black. The unstable trajectories are indicated in different colors in a range from red to blue distributed from the lowest to highest values of $x_{0}$. The unstable region for this value of $C_{j}$ is larger than the unstable regions of higher values of $C_{j}$. From the limits of the variation in the third dimension it is possible to identify the unstable and the stable regions that are clearly separated. The stable region starts from $x_{0}=4.7$ $km$, where the limits of the variation in the third dimension are much lower than in the region before $x_{0}=4.7$ $km$. The red squares indicate the initial conditions whose particles have collided with the asteroid before the trajectory completes $1000$ points in the Poincar\'e surface of section. Comparing the section and the $z$ limits of the variation, it is possible to find chaotic trajectories that produce points in the Poincar\'e surface of section over the stable region, in the same way that it happened in the Poincar\'e surface of section for $C_{j}=2.10$ (Fig. \ref{fig:3d-c}).

\begin{figure*}

\includegraphics[width=17cm,trim = 5mm 40mm 7mm 0mm,clip]{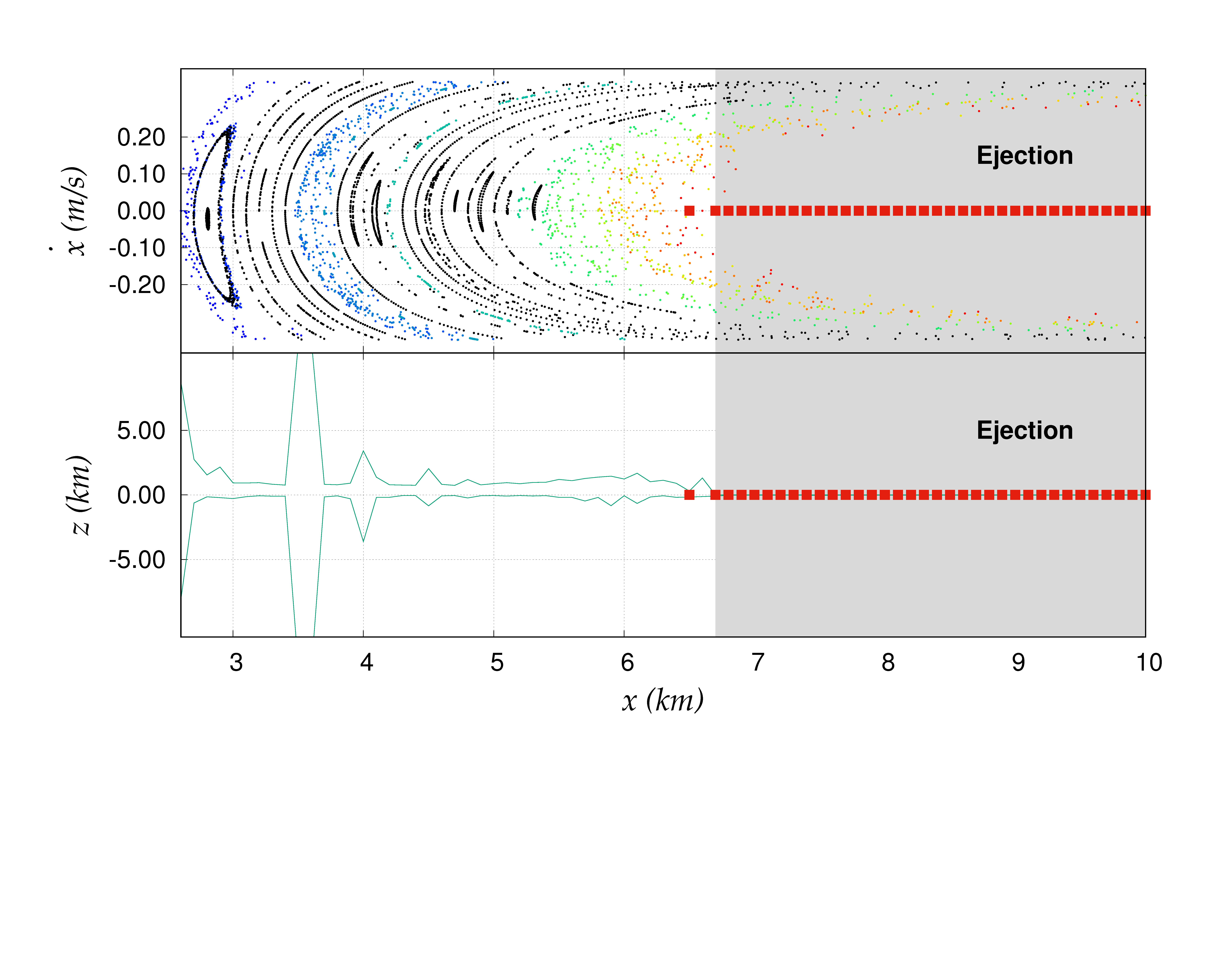}

\caption{Poincar\'e surface of section for $C_{j}=1.20$ and the limits of the variation in the $z$ axis for each initial condition. The chaotic structures are indicated in different colors for each initial condition. Trajectories that ejected from the system are indicated by the red squares.  }
    \label{fig:3d-e}
\end{figure*}

In Fig. \ref{fig:3d-e} the Poincar\'e surface of section for $C_{j}=1.20$ and its respective limits of variation in the third dimension are presented. For this value of $C_{j}$, the forbidden region is widely opened. In the variation within $z$ limits there is not separation between stable and unstable regions but a mix of both. For values higher than $x_{0}=6.7$ $km$, all trajectories are ejected. The red squares indicate the initial conditions of these ejections.

%%%%%%%%%%%%%%%%%%%%%%%%%%%%%%%%%%%%%%%%%%%%%%%%%%%%%%%%%%%%%%%%%%%%%%%%%%%%%%%%%%%%%%%%%%%%%%%%%%%%%%%%%%%%%%%%%%%%
%%%%%%%%%%%%%%%%%%%%%%%%%%%%%%%%%%%%%%%%%%%%%%%%%%%%%%%%%%%%%%%%%%%%%%%%%%%%%%%%%%%%%%%%%%%%%%%%%%%%%%%%%%%%%%%%%%%%
%%%%%%%%%%%%%%%%%%%%%%%%%%%%%%%%%%%%%%%%%%%%%%%%%%%%%%%%%%%%%%%%%%%%%%%%%%%%%%%%%%%%%%%%%%%%%%%%%%%%%%%%%%%%%%%%%%%%
\section{Final comments}

We have shown that the Poincar\'e surface of section technique is applicable to a three-dimension dynamical system of a particle orbiting an irregular body whose gravitational potential is asymmetric. We applied this technique for the asteroid 4179 Toutatis, in order to study stable and unstable regions close to the asteroid. Three families of periodic orbits (projections of \textit{central orbits} in the \textit{rotating plane}) were selected for a detailed analyzes. The first one is a periodic orbits family of first kind, with a very low eccentricity. The other families are of second kind. The second one is a family associated to the resonance $3:1$. The last one is a family associated to the resonance $2:3$. These three families are a sample of the similarity  between a usual two-dimension systems and our approach in the three-dimension case. 

In the evolution of the three periodic orbit families considered we did not identify any kind of bifurcation as those found in the results of \citet{NI}.

We analyze the behavior of a representative set of Poincar\'e surfaces of section combined with the variation in the third dimension. Different structures than the structures of usual two-dimension systems were pointed out in the Poincar\'e surface of section. We found a correlation between the section and the variation in the third dimension. Together with the variation in the third dimension it is possible to identify stable and unstable regions. It is also possible to identify \textit{central orbits} and libration regions around these orbits.

Despite of the three-dimension irregularity of the gravitational potential considered, we have shown that the Poincar\'e Surface of Section allows the identification and location of stable and unstable regions, similarly to what has been done adopting other approaches \citep{Frouard,Araujo2012,2015JIANG}.  

%%%%%%%%%%%%%%%%%%%%%%%%%%%%%%%%%%%%%%%%%%%%%%%%%%%%%%%%%%%%%%%%%%%%%%%%%%%%%%%%%%%%%%%%%%%%%%%%%%%%
%%%%%%%%%%%%%%%%%%%%%%%%%%%%%%%%%%%%%%%%%%%%%%%%%%%%%%%%%%%%%%%%%%%%%%%%%%%%%%%%%%%%%%%%%%%%%%%%%%%%
%%%%%%%%%%%%%%%%%%%%%%%%%%%%%%%%%%%%%%%%%%%%%%%%%%%%%%%%%%%%%%%%%%%%%%%%%%%%%%%%%%%%%%%%%%%%%%%%%%%%
%%%%%%%%%%%%%%%%%%%%%%%%%%%%%%%%%%%%%%%%%%%%%%%%%%%%%%%%%%%%%%%%%%%%%%%%%%%%%%%%%%%%%%%%%%%%%%%%%%%%

\section*{Acknowledgements}

We would like to thank our colleagues Ernesto Vieira Neto, Helton da Silva Gaspar, Luiz Augusto Guimar\~aes Boldrin, Rafael Ribeiro de Sousa and Rodolfo Jos\'e Bueno Rogerio for valuable discussions. This research was supported by the Brazilian agencies CAPES, CNPq and FAPESP(proc. 2016/245610).

%%%%%%%%%%%%%%%%%%%%%%%%%%%%%%%%%%%%%%%%%%%%%%%%%%



\bibliographystyle{mnras}
\bibliography{bilbli} % if your bibtex file is called example.bib

\begin{thebibliography}{}
\makeatletter
\relax
\def\mn@urlcharsother{\let\do\@makeother \do\$\do\&\do\#\do\^\do\_\do\%\do\~}
\def\mn@doi{\begingroup\mn@urlcharsother \@ifnextchar [ {\mn@doi@}
  {\mn@doi@[]}}
\def\mn@doi@[#1]#2{\def\@tempa{#1}\ifx\@tempa\@empty \href
  {http://dx.doi.org/#2} {doi:#2}\else \href {http://dx.doi.org/#2} {#1}\fi
  \endgroup}
\def\mn@eprint#1#2{\mn@eprint@#1:#2::\@nil}
\def\mn@eprint@arXiv#1{\href {http://arxiv.org/abs/#1} {{\tt arXiv:#1}}}
\def\mn@eprint@dblp#1{\href {http://dblp.uni-trier.de/rec/bibtex/#1.xml}
  {dblp:#1}}
\def\mn@eprint@#1:#2:#3:#4\@nil{\def\@tempa {#1}\def\@tempb {#2}\def\@tempc
  {#3}\ifx \@tempc \@empty \let \@tempc \@tempb \let \@tempb \@tempa \fi \ifx
  \@tempb \@empty \def\@tempb {arXiv}\fi \@ifundefined
  {mn@eprint@\@tempb}{\@tempb:\@tempc}{\expandafter \expandafter \csname
  mn@eprint@\@tempb\endcsname \expandafter{\@tempc}}}

\bibitem[\protect\citeauthoryear{{Abell}, {Mazanek}, {Reeves}, {Chodas},
  {Gates}, {Johnson}  \& {Ticker}}{{Abell} et~al.}{2017}]{Abell}
{Abell} P.~A.,  {Mazanek} D.~D.,  {Reeves} D.~M.,  {Chodas} P.~W.,  {Gates}
  M.~M.,  {Johnson} L.~N.,   {Ticker} R.~L.,  2017, in Lunar and Planetary
  Science Conference. p.~2652

\bibitem[\protect\citeauthoryear{Araujo, Winter, Prado  \& Sukhanov}{Araujo
  et~al.}{2012}]{Araujo2012}
Araujo R. A.~N.,  Winter O.~C.,  Prado A. F. B.~A.,   Sukhanov A.,  2012,
  \mn@doi [Monthly Notices of the Royal Astronomical Society]
  {10.1111/j.1365-2966.2012.21101.x}, 423, 3058

\bibitem[\protect\citeauthoryear{Broucke \& Elipe}{Broucke \&
  Elipe}{2005}]{BE05}
Broucke R.,  Elipe A.,  2005, \mn@doi [Regular and Chaotic Dynamics]
  {10.1070/rd2005v010n02abeh000307}, 10, 129

\bibitem[\protect\citeauthoryear{{Bulirsch} \& {Stoer}}{{Bulirsch} \&
  {Stoer}}{1966}]{BS66}
{Bulirsch} R.,  {Stoer} J.,  1966, Journal Numerische Mathematik, 8, 1

\bibitem[\protect\citeauthoryear{{Frouard} \& {Comp{\`e}re}}{{Frouard} \&
  {Comp{\`e}re}}{2012}]{Frouard}
{Frouard} J.,  {Comp{\`e}re} A.,  2012, \mn@doi [\icarus]
  {10.1016/j.icarus.2012.04.026}, \href
  {http://adsabs.harvard.edu/abs/2012Icar..220..149F} {220, 149}

\bibitem[\protect\citeauthoryear{{Geissler}, {Petit}, {Durda}, {Greenberg},
  {Bottke}, {Nolan}  \& {Moore}}{{Geissler} et~al.}{1996}]{G96}
{Geissler} P.,  {Petit} J.-M.,  {Durda} D.~D.,  {Greenberg} R.,  {Bottke} W.,
  {Nolan} M.,   {Moore} J.,  1996, \mn@doi [Icarus] {10.1006/icar.1996.0042},
  \href {http://adsabs.harvard.edu/abs/1996Icar..120..140G} {120, 140}

\bibitem[\protect\citeauthoryear{{H{\'e}non}}{{H{\'e}non}}{1965a}]{henon-a}
{H{\'e}non} M.,  1965a, Annales d'Astrophysique, 28, 499

\bibitem[\protect\citeauthoryear{{H{\'e}non}}{{H{\'e}non}}{1965b}]{henon-b}
{H{\'e}non} M.,  1965b, Annales d'Astrophysique, 28, 992

\bibitem[\protect\citeauthoryear{{H{\'e}non}}{{H{\'e}non}}{1966a}]{henon-d}
{H{\'e}non} M.,  1966a, Annales d'Astrophysique, 1, 49

\bibitem[\protect\citeauthoryear{{H{\'e}non}}{{H{\'e}non}}{1966b}]{henon-c}
{H{\'e}non} M.,  1966b, Annales d'Astrophysique, 1, 57

\bibitem[\protect\citeauthoryear{{H{\'e}non}}{{H{\'e}non}}{1969}]{henon-e}
{H{\'e}non} M.,  1969, Academie des Sciences Paris Comptes Rendus Serie B
  Sciences Physiques, 268, 223

\bibitem[\protect\citeauthoryear{Jefferys}{Jefferys}{1971}]{jefferys}
Jefferys W.,  1971, An Atlas of Surfaces of Section for the Restricted Problem
  of Three Bodies.
Publications of the Department of Astronomy, The University of Texas at
  Austin., Austin, TX, USA.

\bibitem[\protect\citeauthoryear{{Jiang} \& {Baoyin}}{{Jiang} \&
  {Baoyin}}{2016}]{Jiang2016cl}
{Jiang} Y.,  {Baoyin} H.,  2016, \mn@doi [\aj] {10.3847/0004-6256/152/5/137},
  \href {http://adsabs.harvard.edu/abs/2016AJ....152..137J} {152, 137}

\bibitem[\protect\citeauthoryear{Jiang, Yu  \& Baoyin}{Jiang
  et~al.}{2015a}]{Jiang2015cl}
Jiang Y.,  Yu Y.,   Baoyin H.,  2015a, \mn@doi [Nonlinear Dynamics]
  {10.1007/s11071-015-1977-5}, 81, 119

\bibitem[\protect\citeauthoryear{Jiang, Baoyin  \& Li}{Jiang
  et~al.}{2015b}]{2015JIANG}
Jiang Y.,  Baoyin H.,   Li H.,  2015b, \mn@doi [Astrophysics and Space Science]
  {10.1007/s10509-015-2576-0}, 360

\bibitem[\protect\citeauthoryear{Jiang, Baoyin, Wang, Yu, Li, Peng  \&
  Zhang}{Jiang et~al.}{2016}]{JiangSSP}
Jiang Y.,  Baoyin H.,  Wang X.,  Yu Y.,  Li H.,  Peng C.,   Zhang Z.,  2016,
  \mn@doi [Nonlinear Dynamics] {10.1007/s11071-015-2322-8}, 83, 231

\bibitem[\protect\citeauthoryear{{Kawaguchi}, {Fujiwara}  \&
  {Uesugi}}{{Kawaguchi} et~al.}{2008}]{Haya2-b}
{Kawaguchi} J.,  {Fujiwara} A.,   {Uesugi} T.,  2008, \mn@doi [Acta
  Astronautica] {10.1016/j.actaastro.2008.01.028}, \href
  {http://adsabs.harvard.edu/abs/2008AcAau..62..639K} {62, 639}

\bibitem[\protect\citeauthoryear{Kellogg}{Kellogg}{1954}]{Ke54}
Kellogg O.~D.,  1954, Foundations of potential theory.
Springer, Cambridge, MA, USA.

\bibitem[\protect\citeauthoryear{Laplace}{Laplace}{1782}]{La82}
Laplace P.~S.,  1782, Oeuvres de Laplace.
 Vol. 2, Paris, French.

\bibitem[\protect\citeauthoryear{{Lauretta} et~al.,}{{Lauretta}
  et~al.}{2017}]{Osires-b}
{Lauretta} D.~S.,  et~al., 2017, preprint, \href
  {http://adsabs.harvard.edu/abs/2017arXiv170206981L} {}

\bibitem[\protect\citeauthoryear{{Liu}, {Baoyin}  \& {Ma}}{{Liu}
  et~al.}{2011}]{liu11}
{Liu} X.,  {Baoyin} H.,   {Ma} X.,  2011, \mn@doi [Astrophysics and Space
  Science] {10.1007/s10509-011-0732-8}, \href
  {http://adsabs.harvard.edu/abs/2011Ap%26SS.334..357L} {334, 357}

\bibitem[\protect\citeauthoryear{MacMillan}{MacMillan}{1936}]{Mac36}
MacMillan W.~D.,  1936, Dynamics of Rigid Bodies.
New York, USA.

\bibitem[\protect\citeauthoryear{Murray \& Dermott}{Murray \&
  Dermott}{1999}]{murraybook}
Murray C.~D.,  Dermott S.~F.,  1999, Solar System Dynamics.
Cambridge University Press, Cambridge, UK.

\bibitem[\protect\citeauthoryear{{Najid}, {Haj Elourabi}  \&
  {Zegoumou}}{{Najid} et~al.}{2011}]{Naj11}
{Najid} N.-E.,  {Haj Elourabi} E.,   {Zegoumou} M.,  2011, \mn@doi [Research in
  Astronomy and Astrophysics] {10.1088/1674-4527/11/3/008}, \href
  {http://adsabs.harvard.edu/abs/2011RAA....11..345N} {11, 345}

\bibitem[\protect\citeauthoryear{{Neese}}{{Neese}}{2004}]{radar}
{Neese} C.,  2004, NASA Planetary Data System, \href
  {http://adsabs.harvard.edu/abs/2004PDSS...16.....N} {16}

\bibitem[\protect\citeauthoryear{Ni, Jiang  \& Baoyin}{Ni et~al.}{2016}]{NI}
Ni Y.,  Jiang Y.,   Baoyin H.,  2016, \mn@doi [Astrophysics and Space Science]
  {10.1007/s10509-016-2756-6}, 361

\bibitem[\protect\citeauthoryear{{Ostro} et~al.,}{{Ostro} et~al.}{1995}]{Ost95}
{Ostro} S.~J.,  et~al., 1995, \mn@doi [Science] {10.1126/science.270.5233.80},
  \href {http://adsabs.harvard.edu/abs/1995Sci...270...80O} {270, 80}

\bibitem[\protect\citeauthoryear{Poincar{\'e}}{Poincar{\'e}}{1899}]{poincar}
Poincar{\'e} H.,  1899, Les m{\'e}thodes nouvelles de la m{\'e}canique
  c{\'e}leste.
No.~v. 3 in Les m{\'e}thodes nouvelles de la m{\'e}canique c{\'e}leste,
  Gauthier-Villars et fils, Paris, French.

\bibitem[\protect\citeauthoryear{{Prockter}, {Murchie}, {Cheng}, {Klimigis},
  {Farquhar}  \& {Santo}}{{Prockter} et~al.}{2002}]{NShoemaker}
{Prockter} L.,  {Murchie} S.,  {Cheng} A.,  {Klimigis} S.,  {Farquhar} R.,
  {Santo} A.,  2002, Acta Astronautica, 51, 491

\bibitem[\protect\citeauthoryear{{Roll}, {Witte}  \& {Arnold}}{{Roll}
  et~al.}{2016}]{rosetta}
{Roll} R.,  {Witte} L.,   {Arnold} W.,  2016, \mn@doi [Icarus]
  {10.1016/j.icarus.2016.07.004}, \href
  {http://adsabs.harvard.edu/abs/2016Icar..280..359R} {280, 359}

\bibitem[\protect\citeauthoryear{{Rossi}, {Marzari}  \& {Farinella}}{{Rossi}
  et~al.}{1999}]{rossi99}
{Rossi} A.,  {Marzari} F.,   {Farinella} P.,  1999, \mn@doi [Earth, Planets,
  and Space] {10.1186/BF03351592}, \href
  {http://adsabs.harvard.edu/abs/1999EP%26S...51.1173R} {51, 1173}

\bibitem[\protect\citeauthoryear{{Scheeres}, {Ostro}, {Hudson}  \&
  {Werner}}{{Scheeres} et~al.}{1996}]{Sh96}
{Scheeres} D.~J.,  {Ostro} S.~J.,  {Hudson} R.~S.,   {Werner} R.~A.,  1996,
  \mn@doi [Icarus] {10.1006/icar.1996.0072}, \href
  {http://adsabs.harvard.edu/abs/1996Icar..121...67S} {121, 67}

\bibitem[\protect\citeauthoryear{{Scheeres}, {Ostro}, {Hudson}, {DeJong}  \&
  {Suzuki}}{{Scheeres} et~al.}{1998}]{Sh98}
{Scheeres} D.~J.,  {Ostro} S.~J.,  {Hudson} R.~S.,  {DeJong} E.~M.,   {Suzuki}
  S.,  1998, \mn@doi [Icarus] {10.1006/icar.1997.5870}, \href
  {http://adsabs.harvard.edu/abs/1998Icar..132...53S} {132, 53}

\bibitem[\protect\citeauthoryear{{Shustov}, {Naroenkov}  \&
  {Efremova}}{{Shustov} et~al.}{2017}]{Shustov}
{Shustov} B.~M.,  {Naroenkov} S.~A.,   {Efremova} E.~V.,  2017, \mn@doi [Solar
  System Research] {10.1134/S0038094617010038}, \href
  {http://adsabs.harvard.edu/abs/2017SoSyR..51...38S} {51, 38}

\bibitem[\protect\citeauthoryear{{Silva}, {Winter}  \& {Prado}}{{Silva}
  et~al.}{2009}]{silva09}
{Silva} A.~A.,  {Winter} O.~C.,   {Prado} A. F. B.~A.,  2009, \mn@doi
  [Mathematical Problems in Engineering] {10.1155/2009/396267}, 2009

\bibitem[\protect\citeauthoryear{{Venditti}}{{Venditti}}{2013}]{venditti}
{Venditti} F. C.~F.,  2013, PhD thesis, INPE - S\~ao Jos\'e dos Campos - Brazil

\bibitem[\protect\citeauthoryear{{Werner}}{{Werner}}{1994}]{W94}
{Werner} R.~A.,  1994, \mn@doi [Celestial Mechanics and Dynamical Astronomy]
  {10.1007/BF00692875}, \href
  {http://adsabs.harvard.edu/abs/1994CeMDA..59..253W} {59, 253}

\bibitem[\protect\citeauthoryear{{Werner} \& {Scheeres}}{{Werner} \&
  {Scheeres}}{1996}]{WS96}
{Werner} R.~A.,  {Scheeres} D.~J.,  1996, \mn@doi [Celestial Mechanics and
  Dynamical Astronomy] {10.1007/BF00053511}, \href
  {http://adsabs.harvard.edu/abs/1996CeMDA..65..313W} {65, 313}

\bibitem[\protect\citeauthoryear{{Winter}}{{Winter}}{2000}]{w00}
{Winter} O.~C.,  2000, \mn@doi [planss] {10.1016/S0032-0633(99)00082-3}, \href
  {http://adsabs.harvard.edu/abs/2000P%26SS...48...23W} {48, 23}

\bibitem[\protect\citeauthoryear{Winter \& Murray}{Winter \&
  Murray}{1994a}]{atlas1}
Winter O.,  Murray C.,  1994a, Atlas of the Planar, Circular, Restricted
  Three-body Problem: Internal orbits.
 QMW maths notes Vol. 16, Queen Mary and Westfield College, London, UK.

\bibitem[\protect\citeauthoryear{Winter \& Murray}{Winter \&
  Murray}{1994b}]{atlas2}
Winter O.,  Murray C.,  1994b, Atlas of the Planar, Circular, Restricted
  Three-body Problem: External orbits.
 QMW maths notes Vol. 17, Queen Mary and Westfield College, London, UK.

\bibitem[\protect\citeauthoryear{{Winter} \& {Murray}}{{Winter} \&
  {Murray}}{1997a}]{WM97a}
{Winter} O.~C.,  {Murray} C.~D.,  1997a, Astronomy and Astrophysics, \href
  {http://adsabs.harvard.edu/abs/1997A%26A...319..290W} {319, 290}

\bibitem[\protect\citeauthoryear{{Winter} \& {Murray}}{{Winter} \&
  {Murray}}{1997b}]{WM97b}
{Winter} O.~C.,  {Murray} C.~D.,  1997b, Astronomy and Astrophysics, \href
  {http://adsabs.harvard.edu/abs/1997A%26A...328..399W} {328, 399}

\bibitem[\protect\citeauthoryear{Yu, Baoyin  \& Jiang}{Yu et~al.}{2015}]{YBJ}
Yu Y.,  Baoyin H.,   Jiang Y.,  2015, \mn@doi [Monthly Notices of the Royal
  Astronomical Society] {10.1093/mnras/stv1784}, 453, 3270

\makeatother
\end{thebibliography}

%\begin{thebibliography}{99}
%\end{thebibliography}

%%%%%%%%%%%%%%%%%%%%%%%%%%%%%%%%%%%%%%%%%%%%%%%%%%

%%%%%%%%%%%%%%%%%%%%%%%%%%%%%%%%%%%%%%%%%%%%%%%%%%%%%%%%%%%%%%%%%%%%%%%%%%%%%%%%%%%%%%%%%%%%%%%%%%%%%%%%%%%%%%%%%%%%%%%%%%%%%%%%%%%%%%%%%%%%
%%%%%%%%%%%%%%%%%%%%%%%%%%%%%%%%%%%%%%%%%%%%%%%%%%%%%%%%%%%%%%%%%%%%%%%%%%%%%%%%%%%%%%%%%%%%%%%%%%%%%%%%%%%%%%%%%%%%%%%%%%%%%%%%%%%%%%%%%%%%
%%%%%%%%%%%%%%%%%%%%%%%%%%%%%%%%%%%%%%%%%%%%%%%%%%%%%%%%%%%%%%%%%%%%%%%%%%%%%%%%%%%%%%%%%%%%%%%%%%%%%%%%%%%%%%%%%%%%%%%%%%%%%%%%%%%%%%%%%%%%
%%%%%%%%%%%%%%%%%%%%%%%%%%%%%%%%%%%%%%%%%%%%%%%%%%%%%%%%%%%%%%%%%%%%%%%%%%%%%%%%%%%%%%%%%%%%%%%%%%%%%%%%%%%%%%%%%%%%%%%%%%%%%%%%%%%%%%%%%%%%
%%%%%%%%%%%%%%%%%%%%%%%%%%%%%%%%%%%%%%%%%%%%%%%%%%%%%%%%%%%%%%%%%%%%%%%%%%%%%%%%%%%%%%%%%%%%%%%%%%%%%%%%%%%%%%%%%%%%%%%%%%%%%%%%%%%%%%%%%%%%
%%%%%%%%%%%%%%%%%%%%%%%%%%%%%%%%%%%%%%%%%%%%%%%%%%%%%%%%%%%%%%%%%%%%%%%%%%%%%%%%%%%%%%%%%%%%%%%%%%%%%%%%%%%%%%%%%%%%%%%%%%%%%%%%%%%%%%%%%%%%

%---------------------------------------------------------------------------------
%---------------------------------------------------------------------------------

%---------------------------------------------------------------------------------
%---------------------------------------------------------------------------------

%---------------------------------------------------------------------------------
%---------------------------------------------------------------------------------
%---------------------------------------------------------------------------------
%%%%%%%%%%%%%%%%%%%%%%%%%%%%%%%%%%%%%%%%%%%%%%%%%%%%%%%%%%%%%%%%%%%%%%%%%%%%%%%%%%%%%%%%%%%%%%%%%%%%%%%%%%%%%%%%%%%%%%%%%%%%%%%%%%%%%%%%%%%%
%%%%%%%%%%%%%%%%%%%%%%%%%%%%%%%%%%%%%%%%%%%%%%%%%%%%%%%%%%%%%%%%%%%%%%%%%%%%%%%%%%%%%%%%%%%%%%%%%%%%%%%%%%%%%%%%%%%%%%%%%%%%%%%%%%%%%%%%%%%%
%---------------------------------------------------------------------------------ok
%---------------------------------------------------------------------------------ok

%---------------------------------------------------------------------------------ok

%---------------------------------------------------------------------------------ok

%---------------------------------------------------------------------------------
%---------------------------------------------------------------------------------
%---------------------------------------------------------------------------------
%---------------------------------------------------------------------------------

%---------------------------------------------------------------------------------
%%%%%%%%%%%%%%%%%%%%%%%%%%%%%%%%%%%%%%%%%%%%%%%%%%%%%%%%%%%%%%%%%%%%%%%%%%%%%%%%%%%%%%%%%%%%%%%%%%%%%%%%%%%%%%%%%%%%%%%%%%%%%%%%%%%%%%%%%%%%
%%%%%%%%%%%%%%%%%%%%%%%%%%%%%%%%%%%%%%%%%%%%%%%%%%%%%%%%%%%%%%%%%%%%%%%%%%%%%%%%%%%%%%%%%%%%%%%%%%%%%%%%%%%%%%%%%%%%%%%%%%%%%%%%%%%%%%%%%%%%

%---------------------------------------------------------------------------------

%---------------------------------------------------------------------------------
%---------------------------------------------------------------------------------

%---------------------------------------------------------------------------------

%---------------------------------------------------------------------------------

%---------------------------------------------------------------------------------

%---------------------------------------------------------------------------------

%---------------------------------------------------------------------------------
%%%%%%%%%%%%%%%%%%%%%%%%%%%%%%%%%%%%%%%%%%%%%%%%%%%%%%%%%%%%%%%%%%%%%%%%%%%%%%%%%%%%%%%%%%%%%%%%%%%%%%%%%%%%%%%%%%%%%%%%%%%%%%%%%%%%%%%%%%%%

%%%%%%%%%%%%%%%%%%%%%%%%%%%%%%%%%%%%%%%%%%%%%%%%%%%%%%%%%%%%%%%%%%%%%%%%%%%%%%%%%%%%%%%%%%%%%%%%%%%%%%%%%%%%%%%%%%%%
%----------------------------------------------------------------------------------------------------------------------------------------
%trim={<left> <lower> <right> <upper>}
%----------------------------------------------------------------------------------------------------------------------------------------

%trim={<left> <lower> <right> <upper>}
%----------------------------------------------------------------------------------------------------------------------------------------

%trim={<left> <lower> <right> <upper>}
%----------------------------------------------------------------------------------------------------------------------------------------

%trim={<left> <lower> <right> <upper>}
%----------------------------------------------------------------------------------------------------------------------------------------

%trim={<left> <lower> <right> <upper>}
%----------------------------------------------------------------------------------------------------------------------------------------
%%%%%%%%%%%%%%%%%%%%%%%%%%%%%%%%%%%%%%%%%%%%%%%%%%%%%%%%%%%%%%%%%%%%%%%%%%%%%%%%%%%%%%%%%%%%%%%%%%%%%%%%%%%%%%%%%%%%

%%%%%%%%%%%%%%%%% APPENDICES %%%%%%%%%%%%%%%%%%%%%

%%%%%%%%%%%%%%%%%%%%%%%%%%%%%%%%%%%%%%%%%%%%%%%%%%

% Don't change these lines
\bsp	% typesetting comment
\label{lastpage}
\end{document}